\documentclass[aps, reprint,10pt,twocolumn,superscriptaddress]{revtex4-1}
\usepackage{amsmath}
\usepackage{graphicx}
\usepackage{amsfonts}
\usepackage{color}
\usepackage{upgreek}
\usepackage{mathtools}
\usepackage{hyperref}
\usepackage{comment}
\hypersetup{
    colorlinks = true,
    linkcolor =blue,
	citecolor=blue, 
	urlcolor=blue 
}

\begin{document}
\title{Nonclassical radiation from a nonlinear oscillator driven solely by classical  $1/f$ noise}

\author{Archak Purkayastha}
\email{archak.p@phy.iith.ac.in}
\affiliation{Department of Physics, Indian Institute of Technology, Hyderabad 502284, India}
\affiliation{Centre for complex quantum systems, Aarhus University, Nordre Ringgade 1, 8000 Aarhus C, Denmark}

\author{Klaus M\o{}lmer}
\email{klaus.molmer@nbi.ku.dk}
\affiliation{Niels Bohr Insitute, Blegdamsvej 17, 2100 Copenhagen, Denmark}

\date{\today} 

\begin{abstract}
Low-frequency classical $1/f$-noise and quantum noise from low-temperature phonon modes are ubiquitous across various experimental platforms, and are usually considered a hindrance for quantum technological applications. Here we show that the simultaneous action of classical $1/f$ noise and a low-temperature phonon bath on a nonlinear oscillator can result in the generation of nonclassical antibunched radiation without the need for any additional drive. The $1/f$ noise itself provides the source of energy for generation of photons, while the phonon bath prevents heating up to infinite temperature and takes the nonlinear oscillator to a noise-averaged non-equilibrium steady state. The photon current in this non-equilibrium steady state may be detected by a standard wide-band detector. For sufficient nonlinearity and frequency dependence of the effective noise spectrum, the detected radiation can be antibunched. This opens the possibility to turn two of the most ubiquitous intrinsic noises in experimental platforms from  a hindrance to a resource. It shows that wasteful heat from unavoidable noises can be converted into useful radiation. These results are based on the Redfield equation, which provides a rigorously derived general approach to treat any type of weak noise in a quantum system, specified only via the noise spectral function, as we discuss in detail. 
\end{abstract}

\maketitle

\section{ Introduction}
As technology brings us close to realization of practical quantum devices, various sources of noise pose a critical problem. Considerable efforts are employed to characterize and understand various types of noises \cite{Devoret_1997, Clerk_2010, You_2021, Groszkowski_2023, Paladino_2014, Braumuller_2020, Kuhlmann_2013, Yoshihara_2006, Quintana_2017, Norris_2016,  OMalley_2015,Alvarez_2011,Bylander_2011, Yoshihara_2006} and develop protocols to protect against them \cite{Shor_1995, Viola_1999, Rabitz_2000, Faoro_2004,  Yang_2019, Sourza_2012, Malinowski_2017, Baum_2021, Ezzell_2023}.  A possibility only  rarely explored is that the state of the system in presence of such noises might itself be a non-trivial one with potential for technological applications \cite{Purkayastha_2020, Guarnieri_2018,Ghosh_2021,Cao_2021,Aamir_2023,Wang_2023}. Such an alternative approach turns the sources of noise from a hindrance to a resource. Two of the most ubiquitous intrinsic noises that occur across various experimental platforms are the classical $1/f$-noise occurring at low frequencies \cite{Paladino_2014,Braumuller_2020,Kuhlmann_2013,Yang_2019,Yoshihara_2006,
Faoro_2004,Safavi-Naini_2011,Sedlacek_2018,Noel_2019,Prajapati_2021,Jing_2020,Ding_2020,
Carter_2013,Shirotori_2021,Hou_2022,Mercade_2021,Shandilya_2019,Zhang_2023}  and the unavoidable low-temperature quantum noise due to the presence of phonon modes \cite{DeVoe_2002,Porras_2004,Weber_2010,Safavi-Naini_2011,Brownnutt_2015,Hartke_2018,Gullans_2018,Catelani_2019,Wilen_2021,
Siddiqi_2021,Iaia_2022,Bargerbos_2023,Freitag_2017,Mazza_2020,Teoh_2021,
Romaszko_2020,Goel_2021}. In this manuscript, we show that, simultaneous action  of both these noises on a nonlinear oscillator can lead to generation of nonclassical antibunched radiation, without requiring any additional drive. The energy for generating  the radiation is provided by the classical $1/f$-noise, while the coupling to the phonon modes prevents heating and aids antibunching. This provides an intuitive way to filter classical $1/f$-noise into nonclassical radiation. Since nonclassical radiation can aid in applications such as quantum sensing, metrology and imaging \cite{Clark_2021,Ruo-Berchera_2020,Tan_2019,Yadin_2018,Moreau_2017,Brida_2010,Rivas_2010},  it shows that intrinsic heating from unavoidable noises can be utilized to generate useful radiation, perfectly in line with the broad goal of quantum thermal machines \cite{Bhattacharjee_2021,Myers_2022,Cangemi_2023}.

Our work is motivated by the seminal experiment, Ref.~\cite{Quintana_2017}. In this experiment, noise in a flux qubit was studied as a function of frequency over a very wide range, from $0.1$ mHz to $10$ GHz. The symmetric and the anti-symmetric parts of the noise spectral function were experimentally measured. For classical noise, the noise correlation function is real, which ensures that the noise spectral function is symmetric. For noise from a quantum thermal bath, the noise correlation function can be complex,  and its antisymmetric part  can be non-zero. At low temperatures, when thermal excitations are suppressed, the antisymmetric part and the symmetric part become almost equal, which is the signature of solely quantum noise \cite{Clerk_2010,Devoret_1997}. In the experiment, at low frequencies, the symmetric part was seen to have a frequency dependence $\omega^{-\alpha}$ with $\alpha\sim 1$, while the antisymmetric part was a negligible constant. These are the characteristics of classical $1/f$-noise. At high frequency, the symmetric and antisymmetric parts of the noise spectral function became identical, which is the characteristic of purely quantum behavior. The classical to quantum crossover was seen in the noise spectrum at $\sim 1$ GHz.  The frequency dependence  of the quantum noise was found to be $\omega^{s}$ with $s>1$. Such super-ohmic frequency dependencies are often associated with a phonon bath \cite{Weber_2010,Hartke_2018,Gullans_2018}. This experiment motivates the present work investigating the physics near the crossover region, where classical and quantum noises are simultaneously present.

To treat simultaneous presence of classical and quantum noises in a quantum system, we use the weak-coupling Redfield quantum master equation (RE) \cite{Bloch_1957, Redfield_1957}. We discuss how the RE gives a general and controlled way to describe noises in quantum systems, giving access to both the noise averaged expectation values and the two-time correlation functions. We provide rigorous derivations, clearly laying out its accuracy and validity regimes. This description can treat any type of weak noise, specified by its spectral function, without requiring further microscopic modelling. This is crucial because microscopic models for generating $1/f$ noise remains an open problem \cite{Safavi-Naini_2011,Paladino_2014, You_2021}. Despite this, the RE approach gives a consistent way to analyze the physical effects of such noises in quantum systems.

We apply this general framework to explore effect of noises in a nonlinear oscillator. We obtain several analytical results for arbitrary nonlinearity treated within rotating wave approximation. We explore in detail the case of a Kerr nonlinear oscillator in simultaneous presence of classical $1/f$ noise and phonons, showing the generation of antibunched radiation from such intrinsic noises. Finally, we discuss the necessary conditions required to generate antibunched radiation from noises. 

The paper is arranged as follows. In Sec.~\ref{Sec:general_formalism} we give the general formalism to treat classical and quantum noises. In Sec.~\ref{Sec:Nonlinear oscillator}, we apply the formalism to a nonlinear oscillator. This includes both the analytical discussion for arbitrary nonlinearity and the numerical exploration for the Kerr nonlinear oscillator. In Sec.~\ref{Sec:summary_and_outlook}, we give the summary and outlook, highlighting the significance of our work. This is followed by an Appendix, which contains the steps for the rigorous derivations.

\section{ Modelling simultaneous classical and quantum noises}
\label{Sec:general_formalism}

\subsection{The Redfield equation} 
In order to investigate the effect of simultaneous presence of both classical and quantum noises, we need a formalism that allows treatment of both on the same footing. This can be done with  the RE. Consider a system governed by a Hamiltonian $\hat{H}_S$, with small amounts of noise appearing in unwanted couplings of a system observable $\hat{S}$ to sources which are not directly controllable. We call these sources the `baths'. This can be formally written as
\begin{align}
\hat{H} = \hat{H}_S + \epsilon \hat{S} \sum_\ell  \hat{B}_\ell + \sum_\ell \hat{H}_{B_\ell}, 
\end{align}
where $\epsilon \ll 1$ is a dimensionless number giving the strength of coupling to sources of noise, $\hat{B}_\ell$ is the Hermitian operator of the $\ell$th bath coupling to the system, and $\hat{H}_{B_\ell}$ is the Hamiltonian of the $\ell$th bath. We call $\hat{B}_\ell$ the noise operators. We assume that initially, there was no correlation between the system and the baths, and among the baths. This can be formally written as 
\begin{align}
\hat{\rho}_{\rm tot}(0)=\hat{\rho}(0) \prod_{\ell} \hat{\rho}_{B_\ell}(0),
\end{align}
 where $\hat{\rho}_{\rm tot}(0)$ is the initial state of the whole set-up of the system and the baths, $\hat{\rho}(0)$ is the initial state of the system, and $\hat{\rho}_{B_\ell}(0)$ is the initial state of the $\ell$th bath. The initial state of the baths are such that 
\begin{align}
\langle \hat{B}_\ell \rangle_{B} =0,
\end{align}
  where $\langle \ldots \rangle_B={\rm Tr}\left(\ldots \prod_{\ell} \hat{\rho}_{B_\ell}(0)\right) $ denotes trace over bath degrees of freedom. Thus, on average, the noise is zero. The noise correlation functions, i.e, two-time correlation of the bath operators coupling to the system, can  be written as
\begin{align}
\label{noise_correlations}
& \left\langle \hat{B}_\ell(t) \hat{B}_m(0) \right\rangle_{B} = \delta_{\ell m}\int_{-\infty}^{\infty} \frac{d\omega}{2\pi} W^{(\ell)}(\omega)e^{-i\omega t}\nonumber \\
&= \delta_{\ell m}\int_{0}^\infty \frac{d\omega}{\pi} \left[ W_{S}^{(\ell)}(\omega)\cos(\omega t)-iW_{A}^{(\ell)}(\omega)\sin(\omega t) \right], 
\end{align}
where $W^{(\ell)}(\omega)$ is the Fourier transform of the noise correlation function, which is termed the noise spectral function, and
\begin{align}
& W_{S}^{(\ell)}(\omega)=\frac{W_{(\ell)}(\omega)+W_{(\ell)}(-\omega)}{2}, \nonumber \\
& W_{A}^{(\ell)}(\omega)=\frac{W_{(\ell)}(\omega)-W_{(\ell)}(-\omega)}{2},
\end{align}
are the symmetric and the anti-symmetric parts of the noise spectral function.
 Under these conditions, we can obtain the quantum master equation for the long-time dynamics of the system to the leading order in $\epsilon$ via the standard Born-Markov approximation. This gives the RE \cite{Breuer_book},
\begin{align}
&\frac{\partial \hat{\rho}}{\partial t}=i[\hat{\rho}, \hat{H}_S]- \epsilon^2 \left( [\hat{S},\hat{\widetilde{S}}\hat{\rho}(t)]+ {\rm h.c.}\right), \nonumber \\
&\textrm{where }\hat{\widetilde{S}} = \int_0^\infty dt^\prime \hat{S}(-t^\prime) \sum_\ell \left \langle \hat{B}_\ell(t^\prime) \hat{B}_\ell(0) \right\rangle_{B},
\end{align} 
${\rm h.c.}$ stands for Hermitian conjugate, and $\hat{S}(t)=e^{i\hat{H}_S t} \hat{S} e^{-i\hat{H}_S t}$. The steps for rigorous derivation of the RE are given in Appendix~\ref{microscopic_derivation}.

\subsection{Quantum and classical noises}
For classical noise, the notation ${\rm Tr}( \ldots )_B$ should be regarded as the classical average over several noise realizations. With this identification, classical and quantum noises can be treated via RE on the same footing, and we can consider the simultaneous presence of both.

Since $\hat{B}_\ell$ is Hermitian, the presence of the imaginary part in Eq.\eqref{noise_correlations}, which is odd in time, shows explicitly that $\hat{B}_\ell(t)$ and $\hat{B}_\ell(0)$ do not commute, reflecting their quantum nature. On the contrary, classical noises can be thought to be associated with environment degrees of freedom that  commute with each other, making the imaginary part zero. So we have, as discussed before, \cite{Devoret_1997, Clerk_2010}
\begin{align}
W_{A}^{(\ell)}(\omega)=0 \textrm{ for classical noise}, \nonumber \\
W_{A}^{(\ell)}(\omega)\neq 0 \textrm{ for quantum noise}.
\end{align}
If the noise source is a quantum system in thermal equilibrium with inverse temperature $\beta$, the ratio of  $W_{A}^{(\ell)}(\omega)$ to $W_{S}^{(\ell)}(\omega)$ satisfies the so-called Kubo-Martin-Schwinger (KMS) condition,
\begin{align}
\frac{W_{A}^{(\ell)}(\omega)}{W_{S}^{(\ell)}(\omega)}=\tanh\left(\frac{\beta \omega}{2}\right), \textrm{ for quantum thermal noise}.
\end{align}
So, we see that, classical source of noise can be thought of as a quantum thermal bath in the infinite temperature limit, $\beta \rightarrow 0$. In the opposite limit of effectively zero temperature, $\beta \rightarrow \infty$, $\tanh\left(\frac{\beta \omega}{2}\right)=1$, so $W_{A}^{(\ell)}(\omega)={W_{S}^{(\ell)}(\omega)}$. 
The simultaneous presence of both quantum and classical noises therefore leads to a highly out-of-equilibrium situation, taking the system to a non-equilibrium steady state (NESS) in the long-time limit.

Within the above RE description, the noise from various sources are additive, so that the effective total noise spectral functions can be defined: 
\begin{align}
W_S^{tot}(\omega)=\sum_{\ell} W_S^{(\ell)}(\omega),~~W_A^{tot}(\omega)=\sum_{\ell} W_A^{(\ell)}(\omega).
\end{align}
 In the experiment in Ref.~\cite{Quintana_2017}, these total noise spectral functions were measured for a flux qubit.

\subsection{Accuracy and validity regimes}
When modelling classical and quantum noises as above, one must be careful with the accuracy of the RE. This is even more important because the RE does not generically guarantee complete positivity \cite{Tupkary_2023}. Staying within its validity regimes, carefully noting accuracy of various elements of the density matrix obtained from RE can mitigate problems arising from lack of complete positivity \cite{Purkayastha_2020,Tupkary_2022,Hartmann_2020}. It has been shown that, if a unique NESS is reached, the populations in the eigenbasis of the system Hamiltonian are correct to order $O(\epsilon^0)$ while the coherences are  correct to $O(\epsilon^2)$ \cite{Tupkary_2022,Fleming_2011}. Higher order terms obtained from the RE, in either populations or coherences, cannot be trusted. For a bosonic system, the number of bosons must also be restricted in order for the weak-coupling approach to be applicable. Further, the RE is accurate for time $t\gg \tau_M$, where $\tau_M$ is the memory time of the bath, which satisfies,
\begin{align}
\left|\sum_\ell\left \langle \hat{B}_\ell(t) \hat{B}_\ell(0) \right\rangle_{B} \right|<\epsilon,~~\forall ~~t \geq \tau_M,
\end{align}
  in some chosen energy unit (see Appendix~\ref{microscopic_derivation}).

\subsection{Two-time correlations via regression formula}
The RE gives the noise averaged density matrix of the system. The density matrix allows us to obtain expectation values of operators. But, it does not immediately allow to calculation multi-time correlation functions. This can be done using the quantum regression formula. We will be interested in two-time correlation functions of the form
\begin{align}
&\langle \hat{O}_2(t_1) \hat{O}_4(t_1+\tau) \hat{O}_3(t_1+\tau) \hat{O}_1(t) \rangle \nonumber \\
&= {\rm Tr}\left(\hat{O}_2(t_1) \hat{O}_4(t_1+\tau) \hat{O}_3(t_1+\tau) \hat{O}_1(t) \hat{\rho}_{\rm tot}(0) \right),
\end{align}
 where, $\hat{O}_1$, $\hat{O}_2$, $\hat{O}_3$, $\hat{O}_4$ are four system operators, $\hat{O}_\ell(t)=e^{i\hat{H}t} \hat{O}_\ell e^{-i\hat{H}t}$, and $\tau>0$. 
 Using the cyclic property of the trace, this correlation function can be exactly written as
\begin{align}
&\langle \langle \hat{O}_2(t_1) \hat{O}_4(t_1+\tau) \hat{O}_3(t_1+\tau) \hat{O}_1(t_1) \rangle = {\rm Tr}_S\left[ \hat{O}_4 \hat{O}_3 \hat{\widetilde{\rho}}(\tau) \right],
 \nonumber \\
 & \hat{\widetilde{\rho}}(\tau)=  {\rm Tr}_B \left( e^{-i\hat{H}\tau}  \hat{\widetilde{\rho}}_{\rm tot}(t_1) e^{i\hat{H}\tau}\right), \\
& \hat{\widetilde{\rho}}_{\rm tot}(t_1)= \hat{O}_1 \hat{\rho}_{\rm tot}(t_1) \hat{O}_2 \nonumber.
\end{align}
where ${\rm Tr}_S(\ldots)$ refer to trace over system degrees of freedom. Thus, the two-time correlation function is obtained by taking the trace of 
$\hat{O}_4 \hat{O}_3$ with a modified system density matrix $\hat{\widetilde{\rho}}(\tau)$. This modified system density matrix is obtained by dressing the density matrix of the full set-up at time $t_1$ with $\hat{O}_1$ and $\hat{O}_2$, evolving with the full Hamiltonian for a time $\tau$ and taking trace over the bath degrees of freedom. It can be shown that, when $\tau \gg \tau_M$  \cite{Khan_2023}, up to leading order in system-bath coupling strength, $\hat{\widetilde{\rho}}(\tau)$ can be obtained by solving 
\begin{align}
\frac{\partial \hat{\widetilde{\rho}}}{\partial \tau} & \simeq i[\hat{\widetilde{\rho}}, \hat{H}_S]- \epsilon^2 \left( [\hat{S},\hat{\widetilde{S}}\hat{\widetilde{\rho}}(t)]+ {\rm h.c.}\right),~~\forall~\tau \gg \tau_M,
\end{align}
starting with the initial condition
\begin{align}
\hat{\widetilde{\rho}}(t_1)={\rm Tr}_B \left(\hat{\widetilde{\rho}}_{\rm tot}(t_1)\right)=\hat{O}_1 \hat{\rho}(t_1) \hat{O}_2,
\end{align}
where note that $\hat{\rho}(t_1)$ is the system density matrix at time $t_1$. Thus $\hat{\widetilde{\rho}}(\tau)$, for $\tau \gg \tau_M$, can be obtained by solving the RE starting with an appropriately modified initial condition. From  $\hat{\widetilde{\rho}}(\tau)$, the required two-time correlation function can be obtained. The steps for rigorous derivation of the regression formula are given in Appendix.~\ref{microscopic_derivation}.

In this section, we have discussed the general formalism to treat quantum and classical noises in the same setting, and how both expectation values and two-time correlation functions can be obtained consistently via this approach. A crucial point to note is that this formalism based on RE does not need microscopic modelling of the source of noise. Instead, the knowledge of noise spectral function is enough. This is useful because for some kinds of noise, for example, the $1/f$-noise, the underlying microscopic model is often not known \cite{Safavi-Naini_2011,Paladino_2014, You_2021}. The RE approach allows exploration of the effect of even such noises on quantum systems.
In the next section, we see this in the case of a nonlinear oscillator.

\section{A nonlinear oscillator with noises}
\label{Sec:Nonlinear oscillator}

\subsection{The general model and the NESS}
As a simple model of a nonlinear oscillator, we consider the following Hamiltonian,
\begin{align}
\label{general_non_linearity}
\hat{H}_S = \Omega \hat{n}  +  \chi U(\hat{n}), 
\end{align}
where $\hat{n}=\hat{a}^\dagger \hat{a}$ is the number operator, $U(\hat{n})$ is a function of the number operator  and $\hat{a}$ is the bosonic annihilation operator for the oscillator. This nonlinearity can be thought of arising from a nonlinearity that is function of $\hat{x}$, with $\hat{x}=(\hat{a}+\hat{a}^\dagger)$, after neglecting counter-rotating terms (rotating wave approximation).  Considering such a nonlinear oscillator, as opposed to a qubit,  is experimentally relevant because many qubits \cite{Kjaergaard_2020,Dmitriev_2021} are essentially oscillators with a finite nonlinearity. 

The time evolution in absence of coupling to any bath is given by $\hat{a}(t)=e^{-i\hat{\Omega} t}\hat{a}$, where 
\begin{align}
\label{Omega}
\hat{\Omega}=\Omega+ \chi \left[U(\hat{n}+1)-U(\hat{n})\right].
\end{align}
We consider that the coupling to multiple sources of noise occurs via the operator $\hat{x}$. We can then write down the RE for this setting and neglect the counter-rotating terms to obtain \cite{Purkayastha_2016},
\begin{align}
\label{non_lin_osc_RQME}
\frac{\partial \hat{\rho}}{\partial t}&=i[\hat{\rho}, \hat{H}_S]- \epsilon^2\Big([\hat{a}^\dagger, \left(\hat{F}_+ +i\hat{R}_-\right)\hat{a}\hat{\rho}]\nonumber \\
&+\left[\hat{\rho}\left(\hat{F}_-+i\hat{R}_+\right)\hat{a},\hat{a}^\dagger\right]+\rm{h.c.}\Big)
\end{align}
where $
\hat{F}_{\pm} = [W_{S}^{tot}(\hat{\Omega}) \pm W_{A}^{tot}(\hat{\Omega})]/2$, $\hat{R}_{\pm} = \hat{R}_S \pm \hat{R}_A$, $ \hat{R}_S=\mathcal{P} \int \frac{d\omega}{2 \pi} W_{S}^{tot}(\omega)\left(\frac{1}{\hat{\Omega}+\omega}+\frac{1}{\hat{\Omega}-\omega}\right)$, $\hat{R}_A =\mathcal{P} \int \frac{d\omega}{2 \pi}  W_{A}^{tot}(\omega)\left(\frac{1}{\hat{\Omega}+\omega}-\frac{1}{\hat{\Omega}-\omega}\right)$.
 Remarkably,  this equation can be solved analytically to obtain the NESS density matrix. The population, i.e, the photon number distribution, is given by \cite{Purkayastha_2016} (see Appendix~\ref{finding_NESS}),
\begin{align}
\label{photon_distribution}
\rho_n = \rho_0 \prod_{p=1}^n \frac{ \Big[W_{S}^{tot}(\Omega_{p-1}) -  W_{A}^{tot}(\Omega_{p-1}) \Big] }{\Big[ W_{S}^{tot}(\Omega_{p-1}) + W_{A}^{tot}(\Omega_{p-1})\Big]},
\end{align}  
where $\rho_n = \langle n | \hat{\rho} | n \rangle$, $| n \rangle$ being the number eigenbasis, satisfying $\hat{a}| n \rangle= \sqrt{n} | n-1 \rangle$, $\hat{a}^\dagger| n \rangle= \sqrt{n+1} | n+1 \rangle$, $\Omega_n=\langle n | \hat{\Omega} | n \rangle$ and $\rho_0$ being determined by the normalization $\sum_{n=0}^\infty \rho_n=1$. Within the rotating wave approximation, the coherences in the number basis can be argued to be negligible.

\begin{figure}
\includegraphics[width=0.9\columnwidth]{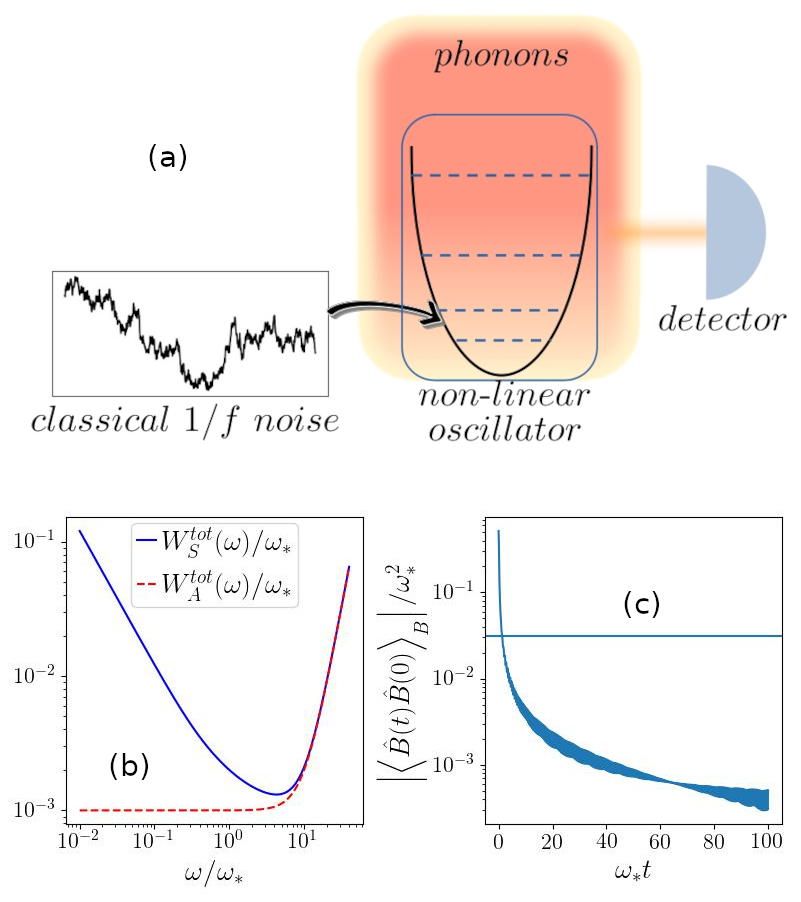}
\caption{\label{fig:schematic} (a) A nonlinear oscillator in simultaneous presence of classical $1/f$ noise, a low temperature super-Ohmic phonon bath and a wide-band photon detector. (b) The total noise spectral functions: $W_S^{tot}(\omega)=W_S^{cl}(\omega)+W_S^{q}(\omega)+W_S^{D}(\omega)$ and $W_A^{tot}(\omega)=W_A^{cl}(\omega)+W_A^{q}(\omega)+W_A^{D}(\omega)$. Parameters: $\beta\omega_*=10$, $s=3$, $\Gamma_{cl}= 10^{-3}$, $\Gamma_q=10^{-6}$, $\Gamma_{D}=10^{-3}$, $\omega_{min}=0.01\omega_*$, $\omega_{max}=50\omega_*$. Here $\Gamma_D$, $\Gamma_{cl}$ and $\Gamma_q$ are dimensionless coupling constants to the detector, the source of classical noise and the phonon bath, respectively. (c) The decay of the total noise correlation function $\left|\left \langle \hat{B}(t) \hat{B}(0) \right\rangle_{B} \right|$ with time. The horizontal line corresponds to $\sqrt{\Gamma_D}$. The value where horizontal line cuts the plot gives the estimate of $\tau_M$. From the plot, we find $\tau_M \sim 2 \omega_*^{-1}$.  }
\end{figure}

From the expression for $\rho_n$, we directly see that if there were only classical sources of noise, i.e, if $W_{A}^{(\ell)}=0~\forall~\ell$, all states would be equally populated. This is true regardless of the strength of nonlinearity $\chi$. This is the infinite temperature state, which, for a bosonic system, is not normalizable and hence ill-defined. This points to a deficiency of the weak-coupling approximation and the relevance of higher order terms, which are important when the system gets highly excited. If, instead, we have only quantum  baths all at the same temperature $\beta$, it can be shown that $\rho_n \propto e^{-\beta [\Omega n + \chi U(n)]}$, which corresponds to the Gibbs state of the system at the temperature of the baths. In presence of both classical and quantum noises, we expect that $\rho_n$ approximates the NESS photon number distribution well, provided the system is  not too excited. This is guaranteed if the temperatures of the quantum thermal baths are low.

\subsection{Classical $1/f$ noise, phonons and detector}
The above description holds for an arbitrary number of noise sources and baths. We now consider the case where the intrinsic noises are classical $1/f$ noise, noise from a phonon bath with super-ohmic spectrum and a wide-band detector, see Fig.~\ref{fig:schematic}(a). For simplicity, we consider that all the baths have a hard  lower cut-off  frequency $\omega_{min}$, and a hard upper cut-off frequency $\omega_{max}$. We take the cut-off frequencies far enough from relevant system frequencies so that their values only  affect the physics negligibly.  The classical $1/f$ noise is described by
\begin{align}
& W_S^{cl}(\omega)=~\Gamma_{cl}~\omega_*\omega^{-1}~\forall~ \omega_{min} \leq \omega \leq \omega_{max}, \nonumber \\
& W_A^{cl}(\omega)=0.
\end{align}
  The phonon bath at inverse temperature $\beta$ is described by 
\begin{align}
& W^{q}_{A}(\omega) = \Gamma_q \frac{\omega^s}{\omega_*^{s-1}},~\forall~\omega_{min} \leq \omega \leq \omega_{max}, \nonumber \\
& W^{q}_{S}(\omega) = W^{q}_{A}(\omega) \coth\left(\frac{\beta \omega}{2}\right).
\end{align}
 A super-ohmic phonon bath corresponds to $s>1$. We choose $s=3$.
The wide-band detector is modelled by a quantum bath with constant spectral function, 
\begin{align}
& W^{D}_{A}(\omega) = \Gamma_D \omega_* ,~\forall~\omega_{min} \leq \omega \leq \omega_{max}, \nonumber \\
& W^{D}_{S}(\omega) = W^{D}_{A}(\omega) \coth\left(\frac{\beta \omega}{2}\right).
\end{align}
 We assume the detector temperature is the  same as that of the phonon bath, and we futher assume the temperature to be low, such that, 
$\beta \Omega \gg 1.$
 The dimensionless parameters $\Gamma_{cl}$, $\Gamma_q$ and $\Gamma_D$ are $O(\epsilon^2)$ and give the strength of coupling to the classical noise, the phonon bath and the detector respectively. The frequency $\omega_*$ provides a scale such that, if  $\Gamma_{cl}=\Gamma_q=\Gamma_D$, $W_S^{(cl)}(\omega_*)=W_A^{(q)}(\omega_*)=W_A^{(D)}(\omega_*)$. In Fig.~\ref{fig:schematic}(b), we plot the total effective noise spectral functions, $W_S^{tot}(\omega)$, $W_A^{tot}(\omega)$, for the chosen set of parameters given in the figure caption. 
For small $\omega\ll 5\omega_*$, $W_S^{tot}(\omega)\sim \omega^{-1}\gg W_A^{tot}(\omega)$, showing that the noise is effectively classical. For $\omega\gg 5\omega_*$, the $W_S^{tot}(\omega)\simeq W_A^{tot}(\omega)\sim \omega^{3}$, so the noise is fully quantum. This is exactly akin to what was experimentally observed in Ref.\cite{Quintana_2017}. 

To apply the RE, we need an estimate of the memory time $\tau_M$ for our chosen bath parameters. In Fig.~\ref{fig:schematic}(c), we do this by plotting $\left|\left \langle \hat{B}(t) \hat{B}(0) \right\rangle_{B} \right|$ with time. For our choice of parameters, the smallest coupling to the bath is the coupling to the detector, $\Gamma_D$, which can be taken as $O(\epsilon^2)$. So $\sqrt{\Gamma_D}\sim O(\epsilon)$, and $\tau_M$ is given by the time after which $\left|\left \langle \hat{B}(t) \hat{B}(0) \right\rangle_{B} \right|<\sqrt{\Gamma_D}$. From the plot, we estimate $\tau_M \sim 2\omega_*^{-1}$.

Since the classical noise acts like an infinite temperature bath, while the phonons and the detector act like low-temperature baths, there is a current of photons excited by the classical noise, that goes into the phonon bath and the detector. The currents can be obtained from Eq.\eqref{non_lin_osc_RQME} by writing the equation for the rate of change of photons in the system as 
\begin{align}
\frac{d \langle \hat{n} \rangle}{dt} = I_{cl} - I_q - I_D,
\end{align}
where $I_{cl}$,  $I_q$ and $I_D$ are obtained by grouping the terms proportional to $\Gamma_{cl}$, $\Gamma_q$ and $\Gamma_D$ respectively, with the corresponding sign. Here, $I_{cl}$ gives the rate of excitation of quanta by the classical noise, $I_q$ gives the rate of dissipation of quanta into the phonon bath, while $I_{D}$ is the photon current into the detector. At NESS,  $\frac{d \langle \hat{n} \rangle}{dt}=0$, so $I_{cl}=I_q + I_D$. Due to its constant spectral function, the photon current into the detector simply becomes $I_D=2\Gamma_D \langle \hat{n} \rangle$, while $I_q$ has a more complicated expression given in terms of the photon distribution (see Appendix~\ref{finding_currents}).

The spectrum of radiation detected by the wide-band detector is given by,
\begin{align}
& S(\omega)={\rm Re}\left[\int_0^\infty d\tau e^{i\omega \tau} g^{(1)}(\tau)\right], \nonumber \\
& \textrm{where } g^{(1)}(\tau)= \lim_{t\to\infty}\left\langle\hat{a}^\dagger(t) \hat{a} (t+\tau) \right\rangle
\end{align}
and ${\rm Re}[\ldots]$ denotes the real part. The counting statistics of the photons is characterized by the second order correlation function 
\begin{align}
g^{(2)}(\tau)=\lim_{t\to\infty}\left[\left\langle\hat{a}^\dagger(t)\hat{a}^\dagger(t~+~\tau) \hat{a}(t+\tau)\hat{a}(t)\right\rangle/\langle\hat{n}(t)\rangle^2\right].
\end{align}
 The photons are antibunched if  $g^{(2)}(0)<g^{(2)}(\tau)$. This is guaranteed to happen if  $g^{(2)}(0)<1$, which corresponds to a sub-Possionian photon distribution \cite{Walls_book}. Since we know the NESS photon distribution,  $g^{(2)}(0)$ can be directly calculated, while $S(\omega)$, $g^{(2)}(\tau)$  can be calculated employing the quantum regression formula, which is applicable for $\tau\gg \tau_M$. Detailed formulas for obtaining $g^{(2)}(\tau)$ and $S(\omega)$ in our setting are given in Appendix~\ref{finding_g2} and Appendix~\ref{finding_spectrum} respectively.
  Note that the above characterization of detected radiation is valid strictly for a low-temperature detector with constant spectral function, as we are considering here.

\subsection{Antibunched radiation from intrinsic noises}

\begin{figure}
\includegraphics[width=\columnwidth]{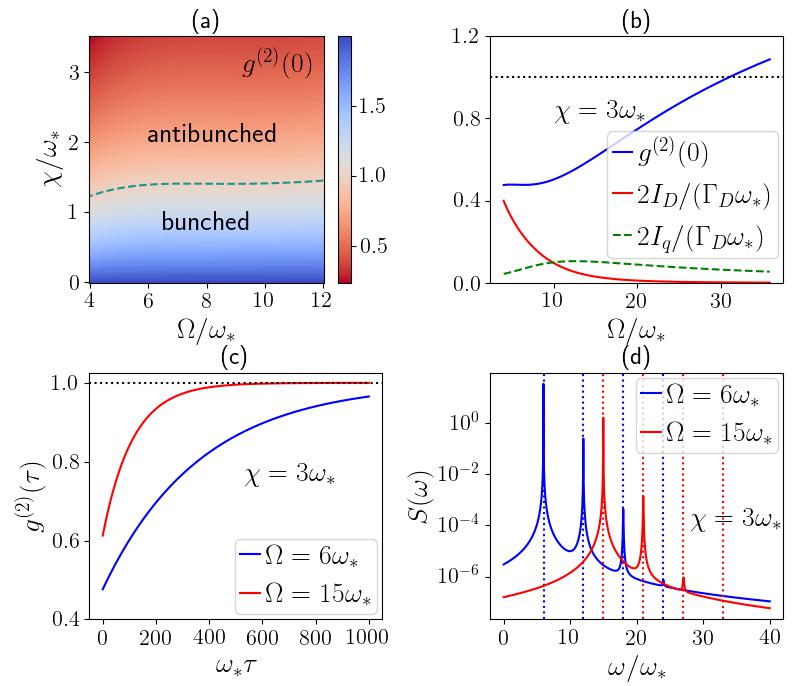}
\caption{\label{fig:g2_currents_spectrum} 
(a) The figure shows $g^{(2)}(0)$ (color-coded), which quantifies the fluctuations in the photon detection signal, as function of $\chi$ and $\Omega$. The dashed line shows the $g^{(2)}(0)=1$ contour.
(b) The rate of dissipation of quanta into the phonon bath ($I_{q}$) and the photon current ($I_D$) into the detector and $g^{(2)}(0)$, are plotted against $\Omega$, for $\chi=3\omega_*$. The currents are multiplied by $2$ to make them visible on the same scale. (c) Steady state $g^{(2)}(\tau)$ for two different values of $\Omega$, with $\chi=3\omega_*$; (d) The frequency spectrum of the photon signal for the same parameters. The vertical dashed lines correspond to $\Omega_n=\Omega+2\chi n$, $n=0,1,2,3$.}
\end{figure}

The above formulation holds for arbitrary choices of nonlinear functions $U(\hat{n})$ (see Eq.\eqref{general_non_linearity}). The nonlinear function only sets $\hat{\Omega}$ (Eq.\eqref{Omega}). We now numerically investigate the experimentally relevant case of Kerr nonlinearity,
\begin{align}
\hat{H}_S = \Omega \hat{n}  +  \chi \hat{n}(\hat{n}-1),
\end{align}
which gives $\hat{\Omega}=\Omega+ 2\chi \hat{n}$, and demonstrate the possibility of antibunched radiation solely driven by the intrinsic $1/f$ noise. We consider $\chi>0$.

In Fig.~\ref{fig:g2_currents_spectrum}(a), we show $g^{(2)}(0)$ as a function of $\chi$ and $\Omega$. We see that for a wide region of parameter space, the photon statistics is sub-Poissionian, which guarantees antibunching. The effective nonlinearity is given by the ratio $\chi/\Omega$. It is interesting to note that sub-Poissionian statistics can occur even with quite small values of $\chi/\Omega$. The Fig.~\ref{fig:g2_currents_spectrum}(b) shows plots of $g^{(2)}(0)$, $I_{D}$, $I_q$ as a function of $\Omega$, keeping $\chi=3\omega_*$. We see that $g^{(2)}(0)$ smoothly crosses over from sub-Poissonian to super-Poissonian values on increasing $\Omega$. This is expected because the effective nonlinearity decreases on increasing $\Omega$. The photon current into the detector, $I_D$, decreases on increasing $\Omega$. This is because smaller $\Omega$ increases coupling to the classical $1/f$ noise, which is the only drive taking the system out-of-equilibrium. The current into the phonon bath, $I_q$, shows a non-monotonic behavior. This arises due to the fact that although the system is farther away from equilibrium at smaller $\Omega$, the coupling to the phonon bath decreases with $\Omega$.

Next, in Fig.~\ref{fig:g2_currents_spectrum}(c), we plot $g^{(2)}(\tau)$ for two chosen parameters. We see that in both cases, $g^{(2)}(\tau)$ monotonically approaches $1$. The monotonic approach may be a peculiarity of the rotating wave approximation, which leads to zero steady state coherence in the number basis. Nevertheless, it is clear that antibunching occurs over a considerable duration of time. Finally, in Fig.~\ref{fig:g2_currents_spectrum}(d), we plot the frequency spectrum of detected radiation for the same two chosen parameters. 
We see that the spectrum has multiple peaks where the higher frequency peaks, corresponding to transitions between the higher excited states, have smaller magnitude. Because of the frequency dependence of the noises, the different peaks also have different widths. 

Overall, the above results clearly demonstrate that it is possible to get nonclassical antibunched radiation from a nonlinear oscillator solely driven by classical $1/f$ noise. Note that, since, for our choice of parameters, the classical to quantum crossover in noise occurs at $\sim 5\omega_*$, the parameters in Fig.~\ref{fig:g2_currents_spectrum} are around the crossover value. If we reduce the parameters to be deep into the classical regime, with finite $\chi$, the photon statistics becomes super-Poissionian again. Contrarily, deep into the quantum regime, $I_D$ goes to zero.

It is useful to consider order of magnitude estimates of various parameters.  Following the experiment in Ref.~\cite{Quintana_2017}, if we set the classical to quantum crossover frequency in noise to be $\sim 1$GHz, the range of $\Omega$ and $\chi$ explored in Fig.~\ref{fig:g2_currents_spectrum}  is then $\sim 0.1$-$10$GHz. The photon current into the detector for such parameters is in the range $1$-$100$kHz.

\subsection{Antibunching without phonons}
\label{without_phonons}

\begin{figure}
\includegraphics[width=\columnwidth]{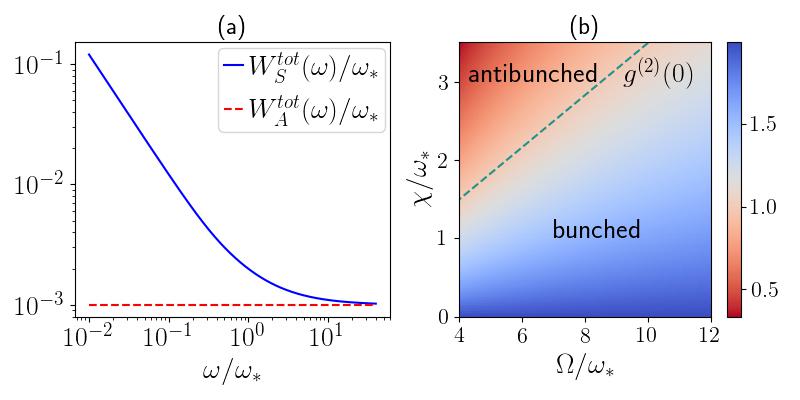}
\caption{\label{fig:no_phonons} (a) The symmetric and antisymmetric parts of total noise spectral functions in absence of the phonon bath.(b) $g^{(2)}(0)$ as a function of $\chi$ and $\Omega$. The dashed line corresponds to $g^{(2)}(0)=1$. Other parameters are the same as in Fig.~\ref{fig:g2_currents_spectrum}. }
\end{figure}

Even without the phonon bath, our set-up has the detector as a low-temperature quantum noise source. In presence of $1/f$ classical noise, we can still work near the crossover region from classical to quantum noise to get antibunched radiation. This is shown in Fig.~\ref{fig:no_phonons}. However, we see that the parameter regime showing antibunching has decreased, as compared to the case in presence of phonons. The presence of the phonon bath therefore aids in antibunching. In experimental situations, it is hard to decouple from phonons. Our results show that coupling to phonons is even advantageous.

\subsection{Necessary requirements for antibunched radiation from noises}

We find that the the minimal necessary properties required to generate antibunched radiation from noises are (a) the noises have different effective temperatures, (b) the system is nonlinear, (c) the noises are frequency dependent. The first condition ensures that the system is out-of-equilibrium, so that there is a photon current.
 To see the necessity of (b) and (c), let us consider cases where either of the following conditions are satisfied: (i) there is no nonlinearity, but there can be frequency dependent classical and quantum noises (ii) the oscillator is nonlinear, but there is no frequency dependence of the noises. The latter condition can arise if there are only white noises, and effectively empty quantum baths with constant spectral functions. In either of the cases, the ratio appearing in the product in Eq.\eqref{photon_distribution} is independent of $n$. Calling this ratio $r$, we find from Eq.\eqref{photon_distribution}, $\rho_n \propto e^{-n\log (r)}$. The NESS in these cases then behaves like that of an effective thermal linear oscillator with effective temperature $\Omega/\log{r}$. While it is interesting that in such cases, even far-from-equilibrium, such an effective temperature can be defined \cite{Kewming_2022,Purkayastha_2016}, it also guarantees that $g^{(2)}(0)=2$ in all such cases. Therefore, both nonlinearity and frequency dependence of noises are required for antibunching.
 
It is to be noted that, phenomenologically modelling dissipation and noise in quantum systems neglecting their frequency dependence is quite popular, owing to the rather simple Lindblad equations that appear due to such approximations. While such approximations are widely used, going beyond them is imperative to uncover interesting and important physics like generation of antibunched radiation from intrinsic noises, as we see above.


\section{Summary and outlook}
\label{Sec:summary_and_outlook}

{\it Utilizing intrinsic noises ---}
A sufficiently nonlinear oscillator can generically produce antibunched radiation when driven. In such a case, the drive would provide the energy required for generating the radiation. The main significance of our work is that, instead of such an external drive, we propose to directly use the intrinsic noises in an experimental setting to generate antibunched light. 
Two ubiquitous sources of noise occurring across various experimental platforms are the low frequency classical $1/f$ noise \cite{Paladino_2014,Braumuller_2020,Kuhlmann_2013,Yang_2019,Yoshihara_2006,
Faoro_2004,Safavi-Naini_2011,Sedlacek_2018,Noel_2019,Prajapati_2021,Jing_2020,Ding_2020,
Carter_2013,Shirotori_2021,Hou_2022,Mercade_2021,Shandilya_2019,Zhang_2023} and the quantum noise due to coupling to phonons \cite{DeVoe_2002,Porras_2004,Weber_2010,Safavi-Naini_2011,Brownnutt_2015,Hartke_2018,Gullans_2018,Catelani_2019,Wilen_2021,
Siddiqi_2021,Iaia_2022,Bargerbos_2023,Freitag_2017,Mazza_2020,Teoh_2021,
Romaszko_2020,Goel_2021}. These sources of noise, which are now quite well characterized, are usually considered a hindrance for quantum technology applications.
We show that simultaneous action of these two noises on a nonlinear oscillator can lead to emission of nonclassical, antibunched radiation, without requiring any further drive. 
The $1/f$ noise itself provides the energy required for generation of photons, while the phonon bath prevents heating and aids antibunching. 

Nonclassical radiation has been shown to aid in quantum sensing, metrology and imaging applications \cite{Clark_2021,Ruo-Berchera_2020,Tan_2019,Yadin_2018,Moreau_2017,Brida_2010,Rivas_2010}. Therefore, our main result turns the intrinsic noises from a hindrance to a resource.  Experimentally accessible parameter regimes, which were previously deemed unfavourable for technological applications due to the intrinsic noises, may now be explored for generation of antibunched radiation. To our knowledge, the possibility to utilize the ubiquitous classical $1/f$ noise was not known before, although some  works go towards utilizing the coupling to phonons 
\cite{Guarnieri_2018,Ge_2019,Purkayastha_2020,Gambetta_2020,Davoudi_2021,
Tao_2022,Mendonca_2023,Slobodeniuk_2022,Slobodeniuk_2023}. This is also consistent with the broad motivation for quantum thermal machines \cite{Bhattacharjee_2021,Myers_2022,Cangemi_2023}, which is to convert wasteful heating in quantum devices (here, form `infinite temperature' classical $1/f$ noise) into useful energy (here, antibunched radiation).

{\it A general framework to treat noises in quantum systems ---}
Apart from the above, we also provide a detailed discussion of how the RE provides a general and rigorous framework to treat all types of noises in a quantum system. The noises are all treated in an equal footing, irrespective of whether they are quantum or classical. This treatment does not require explicit modelling of the sources of the noise, but rather, only the knowledge of the noise spectral function. The approximations required are controlled, and their accuracy and validity regimes are also discussed.  This framework can be envisaged to be useful across many other settings to characterize the noises in quantum devices.

The RE does not generically preserve complete positivity of the density matrix, but often such concerns can be mitigated by carefully staying within its accuracy and validity regimes \cite{Tupkary_2022,Purkayastha_2020,Hartmann_2020}. It can also be reduced to several types of Lindblad equations through further approximations (for example, \cite{Becker_2021, Trushechkin_2021, Davidovic_2020, Nathan_2020, Mozgunov_2020}). These approximations ensure complete positivity at the expense of lower accuracy. Most often, they may also lead to some other fundamental limitations, such as lack of thermalization and violation of local conservation laws \cite{Tupkary_2022,Tupkary_2023}. Alternatively, our description in terms of RE can be improved by employing the recently discovered canonically consistent quantum master equation approach \cite{Becker_2022}.

We would like to mention here that significant effort has been dedicated to microscopically model sources of classical $1/f$-noise \cite{Safavi-Naini_2011,Paladino_2014, You_2021}. While such modelling remains an open question, our framework provides a controlled way to completely avoid such microscopic modelling and still analyze the effects of $1/f$-noise in a realistic setting.

{\it Future directions---}
From above, it is clear that our results should be of interest to researchers across quantum science and technology, solid state physics, quantum optics, thermodynamics and statistical physics. We re-iterate that controlled simultaneous coupling to $1/f$-noise and phonons in a nonlinear oscillator (flux qubit) has already been demonstrated \cite{Quintana_2017}, although the photon statistics of the radiation generated has not yet been measured.  We believe that the observed antibunching in our setting can be improved via further filtering to detect only photons within a small frequency range. Investigations in this direction will be carried out in future works. Other future works  may explore scaling the system to several nonlinear oscillators and going beyond the weak coupling regime.

{\it Acknowledgements ---}
The authors acknowledge support from the Danish National Research Foundation
through the Center of Excellence “CCQ”
(Grant No. DNRF 156).

\appendix

\section*{Appendix}

\section{Microscopic derivation of RE and regression formula}
\label{microscopic_derivation}

\subsection{General microscopic setting}

We go back to the fully microscopic starting point. The full set-up of system and the baths is governed by the Hamiltonian 
\begin{align}
\label{full_H}
\hat{H} = \hat{H}_S + \epsilon \hat{S}\hat{B} + \hat{H}_{B},
\end{align}
 and the initial state of the full set-up is taken as $\hat{\rho}_{\rm tot}(0)=\hat{\rho}(0) \hat{\rho}_{B}(0)$. Here we have defined the composite Hamiltonian of all the baths $\hat{H}_{B}=\sum_\ell \hat{H}_{B_\ell}$, the composite initial state of all the baths $\hat{\rho}_{B}(0)=\prod_\ell \hat{\rho}_{B_\ell}(0)$, and the composite system-bath coupling $\hat{B}=\sum_\ell \hat{B}_\ell$. Without much loss of generality, we assume $\langle \hat{B} \rangle_{B} =0$.  The state of the system at time $t$ is given by
\begin{align}
\hat{\rho}(t) = {\rm Tr}_B\left( e^{-\hat{H}t} \rho_{\rm tot}(0) e^{-\hat{H}t}\right).
\end{align}
We define the following superoperators
\begin{align}
\hat{P}(\bullet)={\rm Tr}_B (\bullet) \hat{\rho}_B,~~\hat{Q}=\hat{\mathbb{I}}-\hat{P},~\hat{\mathcal{L}}=i[\bullet,\hat{H}],
\end{align}
where $\hat{\mathbb{I}}$ is the identity superoperator. In terms of these superoperators, the evolution equation for $\hat{\rho}(t)$ can be exactly written as \cite{Breuer_book}
\begin{align}
\label{exact_QME1}
& \frac{\partial \hat{\rho}}{\partial t}= i[\hat{\rho},\hat{H}_S] + \int_0^t dt^\prime \hat{K}(t^\prime)[\hat{\rho}(t-t^\prime)]  \nonumber \\ 
&  \hat{K}(t)[\bullet] = {\rm Tr}_B \left( \hat{\mathcal{L}} e^{t \hat{Q} \hat{\mathcal{L}}} \hat{Q} \hat{\mathcal{L}}\hat{P}[\bullet] \right).
\end{align}
$\hat{K}(t)$ is called the memory kernel superoperator. The exact same evolution can also be equivalently written for $t_1>0$ as \cite{Breuer_book}
\begin{align}
\label{exact_QME2}
& \frac{\partial \hat{\rho}}{\partial t}= i[\hat{\rho},\hat{H}_S] + \int_0^{t-t_1} dt^\prime \hat{K}(t^\prime)[\hat{\rho}(t-t^\prime)]+\hat{\mathcal{I}}(t-t_1)[\hat{\rho}(t_1)]  \nonumber \\ 
&  \hat{\mathcal{I}}(t)[\bullet] = {\rm Tr}_B \left( \hat{\mathcal{L}} e^{t \hat{Q} \hat{\mathcal{L}}} \hat{Q} [\bullet] \right).
\end{align}
The occurrence of the superoperator $\hat{\mathcal{I}}(t)$ stems from the fact that at time $t_1$, the system and the baths are no longer in product state. Note that both the superoperators $\hat{K}(t)$ and $\hat{\mathcal{I}}(t)$ depend on the initial state of the bath $\hat{\rho}_B(0)$ and the full Hamiltonian $\hat{H}$, but are independent of the initial state of the system $\hat{\rho}(0)$.

The above equations are exact. We now perform the Born-Markov approximation to obtain the RE.
In the following, we assume that time has been rescaled to some unit, so that it can be considered as a dimensionless parameter.

\subsection{Born-Markov approximation and RE}
In doing the Born-Markov approximation, first, the memory kernel in Eq.\eqref{exact_QME1} is expanded in powers of system-bath coupling strength $\epsilon$, and the leading order is retained, higher orders are neglected. This gives \cite{Breuer_book,Tupkary_2022}
\begin{align}
\label{finite_time_RE}
& \frac{\partial \hat{\rho}}{\partial t} \simeq i[\hat{\rho},\hat{H}_S] + \epsilon^2 \int_0^t dt^\prime \hat{K}^{(RE)}(t^\prime)[\hat{\rho}(t-t^\prime)], \nonumber \\
&\hat{K}^{(RE)}(t)[\bullet]=- \left( [\hat{S},\hat{{S}}(-t^\prime)\bullet] \left \langle \hat{B}(t^\prime) \hat{B}(0) \right\rangle_{B}+ {\rm h.c.}\right).
\end{align}
Let the bath correlation function decay with time. Then there is a time scale $\tau_M$, such that
\begin{align}
\label{tau_M_def}
\left|\left \langle \hat{B}(t) \hat{B}(0) \right\rangle_{B} \right|<\epsilon,~~\forall ~~t \geq \tau_M.
\end{align}
For $t\gg \tau_M$, setting the upper limit of time integral to infinity in  Eq.\eqref{finite_time_RE} makes only errors in higher order in $\epsilon$. So, we get, 
\begin{align}
\label{RQME}
& \frac{\partial \hat{\rho}}{\partial t} \simeq i[\hat{\rho}, \hat{H}_S]- \epsilon^2 \left( [\hat{S},\hat{\widetilde{S}}\hat{\rho}(t)]+ {\rm h.c.}\right),  \nonumber \\
&  \hat{\widetilde{S}} = \int_0^\infty dt^\prime \hat{S}(-t^\prime) \sum_\ell \left \langle \hat{B}_\ell(t^\prime) \hat{B}_\ell(0) \right\rangle_{B},~~\forall~t\gg \tau_M,
\end{align}
which is the RE that we use to describe quantum and classical noises on the same footing. The above two approximations together are called the Born-Markov approximations.

\subsection{Loss of information about initial system-bath correlations}
\label{loss_of_initial_correlation}

We can now write Eq.\eqref{exact_QME2} also applying the Born-Markov approximation to the memory kernel superoperator. In doing so, we get
\begin{align}
\label{RQME_with_initial_correlations}
\frac{\partial \hat{\rho}}{\partial t} & \simeq i[\hat{\rho}, \hat{H}_S]- \epsilon^2 \left( [\hat{S},\hat{\widetilde{S}}\hat{\rho}(t)]+ {\rm h.c.}\right) \nonumber \\
&+ \hat{\mathcal{I}}(t-t_1)[\hat{\rho}(t_1)],~~\forall~t-t_1 \gg \tau_M.
\end{align}
Since this equation should describe the exact same state as described by Eq.\eqref{RQME}, it follows that we must have
\begin{align}
\left | \left | \hat{\mathcal{I}}(t-t_1)[\hat{\rho}(t_1)] \right | \right | < \epsilon,~~\forall~t-t_1 \gg \tau_M,
\end{align} 
where $\left | \left | \hat{O} \right | \right |$ is the norm of the operator $\hat{O}$, so that the last term in Eq.\eqref{RQME_with_initial_correlations} can be dropped. In other words, the information that the system was correlated with the bath at time $t_1$ is lost in a time $\tau_M$.
Since this should be true for all possible choices of the time $t_1$ and all possible choice of $\hat{\rho}(0)$, this should be a property of the superoperator $\hat{\mathcal{I}}(t)$.

\subsection{Two-time correlations and quantum regression}
\label{General_regression}

As mentioned in the main text, we are interested in correlations function involving four system operators and two times 
\begin{align}
\label{two_time_correlation}
&\langle \hat{O}_2(t_1) \hat{O}_4(t_1+\tau) \hat{O}_3(t_1+\tau) \hat{O}_1(t) \rangle \nonumber \\
&= {\rm Tr}\left(\hat{O}_2(t_1) \hat{O}_4(t_1+\tau) \hat{O}_3(t_1+\tau) \hat{O}_1(t) \hat{\rho}_{\rm tot}(0) \right),
\end{align}
 where $\hat{O}_\ell(t)=e^{i\hat{H}t} \hat{O}_\ell e^{-i\hat{H}t}$, and $\tau>0$, which can be written as
\begin{align}
&\langle \langle \hat{O}_2(t_1) \hat{O}_4(t_1+\tau) \hat{O}_3(t_1+\tau) \hat{O}_1(t_1) \rangle = {\rm Tr}_S\left[ \hat{O}_4 \hat{O}_3 \hat{\widetilde{\rho}}(\tau) \right],
 \nonumber \\
 & \hat{\widetilde{\rho}}(\tau)=  {\rm Tr}_B \left( e^{-i\hat{H}\tau}  \hat{\widetilde{\rho}}_{\rm tot}(t_1) e^{i\hat{H}\tau}\right), \\
& \hat{\widetilde{\rho}}_{\rm tot}(t_1)= \hat{O}_1 \hat{\rho}_{\rm tot}(t_1) \hat{O}_2 \nonumber.
\end{align}
where ${\rm Tr}_S(\ldots)$ refer to trace over system degrees of freedom. The correlation function is therefore obtained by taking the `expectation value' with a generalized system density matrix $\hat{\widetilde{\rho}}(\tau,t_1)$. This generalized system density matrix is obtained taking the full density matrix dressed by the operators $\hat{O}_1$ and $\hat{O}_2$, evolving it with the full Hamiltonian $\hat{H}$ up to time $\tau$ and tracing out the bath degrees of freedom. The operation comprising evolution via the full Hamiltonian and tracing out the bath degrees of freedom is exactly the same as that required in obtaining the system density matrix. Here, however, the `initial state' $\hat{\widetilde{\rho}}_{\rm tot}(t_1)$ has system-bath correlations. So, without any further approximation,the evolution equation for $\hat{\widetilde{\rho}}(\tau) $ with $\tau$ is the same as Eq.\eqref{exact_QME2},
\begin{align}
\label{exact_QME2_regression}
 \frac{\partial \hat{\widetilde{\rho}}}{\partial \tau}= & i[\hat{\widetilde{\rho}},\hat{H}_S] + \int_0^{\tau-t_1} dt^\prime \hat{K}(t^\prime)[\hat{\widetilde{\rho}}(\tau-t^\prime)] \nonumber \\
&+\hat{\mathcal{I}}(\tau-t_1)[\hat{\widetilde{\rho}}(t_1)].
\end{align}
Given the `initial condition'
\begin{align}
\hat{\widetilde{\rho}}(0) = {\rm Tr}_B \left(   \hat{\widetilde{\rho}}_{\rm tot}(t_1) \right)= \hat{O}_1 \hat{\rho}(t_1) \hat{O}_2,
\end{align}
where, note that $\hat{\rho}(t_1)$ is the system density matrix at time $t_1$, we can in principle solve Eq.\eqref{exact_QME2_regression} to obtain $\hat{\widetilde{\rho}}(\tau)$. Then taking the `expectation value' with respect to this generalized system density matrix, we obtain the desired correlation function.

This is the most general form of quantum regression, which holds without any approximation. Note that this has been derived for $\tau>0$, so a notion of time-ordering is already built-in. For $\tau<0$, the analog of the above formalism has to be re-derived, which will change the initial condition and the final operators in the formal expectation value.

\subsection{Quantum regression with Born-Markov approximation}
\label{Regression_with_Born_Markov_approx}

From the discussions in Secs.~\ref{loss_of_initial_correlation} and \ref{General_regression}, it is now easy to see that under weak system-bath coupling, the Born-Markov approximation leads to
\begin{align}
\frac{\partial \hat{\widetilde{\rho}}}{\partial \tau} & \simeq i[\hat{\widetilde{\rho}}, \hat{H}_S]- \epsilon^2 \left( [\hat{S},\hat{\widetilde{S}}\hat{\widetilde{\rho}}(t)]+ {\rm h.c.}\right),~~\forall~\tau \gg \tau_M.
\end{align}
It follows that, to leading order in system-bath coupling, two-time correlations can be obtained by solving the RE with appropriately modified initial conditions and taking the trace after multiplying by appropriate operators.

\section{Finding the NESS population distribution}
\label{finding_NESS}
Here we show the steps to derive Eq.\eqref{photon_distribution} from Eq.\eqref{non_lin_osc_RQME}. Writing the evolution equation for $\rho_n = \langle n | \hat{\rho} | n \rangle$, $| n \rangle$ being the number eigenbasis, satisfying $\hat{a}| n \rangle= \sqrt{n} | n-1 \rangle$, $\hat{a}^\dagger| n \rangle= \sqrt{n+1} | n+1 \rangle$, gives
\begin{align}
\label{photon_number_equation}
\frac{d \rho_n}{dt}= -\epsilon^2 \left[\rho_n (C_n + D_n) - \rho_{n-1} C_{n-1} - \rho_{n+1} D_{n+1} \right].
\end{align} 
Here,
\begin{align}
\label{C_and_D}
& C_n=\sum_{\ell} C_n^{(\ell)},~~D_n = \sum_\ell D_n^{(\ell)} \nonumber \\
& C_n^{(\ell)} = (n+1) \left(W_{S}^{(\ell)}(\Omega_n) - W_{A}^{(\ell)}(\Omega_n)\right) \nonumber \\
& D_n^{(\ell)} = n \left(W_{S}^{(\ell)}(\Omega_{n-1}) + W_{A}^{(\ell)}(\Omega_{n-1})\right),
\end{align}
where $\Omega_n=\Omega+2n\chi$. For NESS, the left hand side of Eq.\eqref{photon_number_equation} is zero. This takes it to the form of a difference equation. The difference equation can be obtained by writing the equation sequentially for various values of $n$ and adding them. This yields
\begin{align}
\label{NESS_soln}
\rho_{n-1} C_{n-1} = \rho_n D_n \Rightarrow \rho_n = \rho_0 \prod_{p=1}^n \frac{C_{p-1}}{D_p}, n=1,2,3,\ldots.
\end{align}
Combining with Eq.\eqref{C_and_D}, we get Eq.\eqref{photon_distribution}.

\section{Finding currents}
\label{finding_currents}
The currents are defined by the continuity equation
\begin{align}
\label{continuity}
\frac{d \langle \hat{n} \rangle}{dt}=-\sum_\ell I_\ell,
\end{align}
where $I_\ell$ is the current into the $\ell$th bath. From Eq.\eqref{photon_number_equation}, after little algebra, we find that
\begin{align}
\label{dndt}
\frac{d \langle \hat{n} \rangle}{dt}=\sum_{n=0}^\infty n \frac{d\rho_n}{dt}=\sum_{n=0}^\infty \rho_n (C_n - D_n).
\end{align}
Combining Eqs.\eqref{C_and_D}, \eqref{continuity} and \eqref{dndt}, we see that current into the $\ell$th bath is given by
\begin{align}
I_\ell = \sum_{n=0}^\infty \rho_n (D_n^{(\ell)} - C_n^{(\ell)}).
\end{align}
Putting the NESS value of $\rho_n$, the NESS currents can be obtained.
For the detector, which is modelled as an empty bath with constant spectral function, the expression simplifies to 
\begin{align}
\label{detector_current}
I_D=2\Gamma_D \langle \hat{n} \rangle,
\end{align}
where $\langle \hat{n} \rangle$ is the average photon number at NESS.
The expression for current into the phonon bath does not allow any major further simplification.  

\section{Obtaining $g^{(2)}(\tau)$}
\label{finding_g2}

The second order correlation function is given by 
\begin{align}
g^{(2)}(\tau)=\langle \hat{a}^\dagger(t) \hat{a}^\dagger (t+\tau) \hat{a}(t+\tau) \hat{a} \rangle/\langle\hat{n}(t)\rangle^2.
\end{align}
The denominator is easily obtained with $\langle\hat{n}(t)\rangle=\sum_{n=0}^\infty n \rho_n(t)$, where $\rho_n(t)$ is the solution of Eq.\eqref{photon_number_equation}. The numerator is exactly of the form of correlation function in Eq.\eqref{two_time_correlation}. Following the discussion in Secs.\ref{General_regression} and \ref{Regression_with_Born_Markov_approx}, we see that 
\begin{align}
\label{g2_trace}
\langle\hat{n}(t)\rangle^2 g^{(2)}(\tau) = {\rm Tr}_S\left[ \hat{n}\hat{\widetilde{\rho}}(\tau) \right],
\end{align}
where, for $\tau \gg \tau_M$, $\hat{\widetilde{\rho}}(\tau)$ is obtained from solving the RE with the initial condition 
\begin{align}
\label{g2_initial_condition}
\hat{\widetilde{\rho}}(0)=\hat{a} \hat{\rho}(t) \hat{a}^\dagger= \sum_{n=1}^\infty n \rho_n(t) | n-1 \rangle \langle n-1 |.
\end{align}
Since the initial condition is diagonal in the number basis and the final trace also requires only the diagonal elements in the same basis, instead of the full RE, we can use Eq.\eqref{photon_number_equation}. Thus, the numerator of $g^{(2)}(\tau)$ is obtained by evolving the initial condition in Eq.\eqref{g2_initial_condition} with Eq.\eqref{photon_number_equation}, and calculating the trace in Eq.\eqref{g2_trace}.

In the main text, we are interested in the NESS, so we take $t\to \infty$. This amounts to using the NESS population distribution given in Eq.\eqref{NESS_soln} in calculating the initial condition and the denominator.

\section{Obtaining the spectrum $S(\omega)$}
\label{finding_spectrum}

The spectrum of radiation detected by the wide-band detector is given by,
\begin{align}
\label{spectrum_def}
& S(\omega)={\rm Re}\left[\int_0^\infty d\tau e^{i\omega \tau} g^{(1)}(\tau)\right], \\
& g^{(1)}(\tau)= \lim_{t\to\infty}\left\langle\hat{a}^\dagger(t) \hat{a} (t+\tau) \right\rangle, 
\end{align}
where ${\rm Re}[\ldots]$ denotes the real part. The spectrum satisfies the sum rule,
\begin{align}
\label{sum_rule}
\int_{\omega_{min}}^{\omega_{max}}  \frac{d\omega}{\pi} S(\omega)= \langle \hat{n} \rangle,
\end{align}
where $\langle \hat{n} \rangle$ is the average photon number at NESS. The relation between $S(\omega)$ and $I_D$ now follows from Eq.\eqref{detector_current}.

Following the discussion in Secs.\ref{General_regression} and \ref{Regression_with_Born_Markov_approx}, we find that
\begin{align}
\label{Pn_def}
g^{(1)}(\tau)={\rm Tr}_S\left[ \hat{a}\hat{\widetilde{\rho}}(\tau) \right]=\sum_{n=1}^\infty P_n(\tau),~P_n(\tau)=\sqrt{n}\widetilde{\rho}_{n,n-1}(\tau),
\end{align}
where $\hat{\widetilde{\rho}}(\tau)=\sum_{n,m=0}^\infty \widetilde{\rho}_{n,m}(\tau) |n\rangle\langle m |$ is obtained by solving the RE with the initial condition
\begin{align}
\label{g1_initial_condition}
\hat{\widetilde{\rho}}(0)=\hat{\rho}\hat{a}^\dagger=\sum_{n=0}^\infty \sqrt{n}\rho_{n} | n\rangle \langle n-1 |,
\end{align}
where $\hat{\rho}=\sum_{n=0}^\infty \rho_n | n\rangle \langle n |$ is the NESS density matrix of the system.

From the RE Eq.\eqref{non_lin_osc_RQME}, we can directly find an evolution equation for $P_n(\tau)$. After some tedious but straightforward algebra, the evolution equation for $P_n(\tau)$ becomes,
\begin{align}
\label{Pn_diff_eqn}
\frac{d P_n}{d\tau} =& -\left[i\Omega_n P_n+ \epsilon^2\left(A_n + K_n +  G^{(1)}_{n-1} - G^{(2)}_{n-1}\right)\right] P_n \nonumber \\
&- \epsilon^2 A_{n+1} P_{n+1} - \epsilon^2 K_{n-1} P_{n-1}, 
\end{align}
with the following definitions
\begin{align}
& A_{n}=(n-1)\left( G^{(1)}_{n-1} + G^{*(1)}_{n-2}\right), \nonumber\\
& K_{n}=(n+1)\left( G^{(2)}_{n-1} + G^{*(2)}_{n}\right), \\
& G_{n}^{(1)}=F_+(\Omega_n) + i R_-(\Omega_n). \nonumber \\
& G_{n}^{(2)}=F_-(\Omega_n) + i R_+(\Omega_n), \nonumber
\end{align}
where $F_\pm(\Omega_n)=\langle n | \hat{F}_\pm | n \rangle$ and $R_\pm(\Omega_n)=\langle n | \hat{R}_\pm | n \rangle$. From Eqs.\eqref{Pn_def} and \eqref{g1_initial_condition}, the initial condition for the above equation becomes
\begin{align}
P_n(0) = n \rho_n,~\forall n\geq 1.
\end{align}
Putting some finite but large enough upper cut-off $n_{max}$ on the number of photons in the system, we can rewrite Eq.\eqref{Pn_diff_eqn} as a matrix equation
\begin{align}
\label{Pn_matrix_equation}
\frac{d P^{vec}}{d\tau}=-M P^{vec},
\end{align}
where $P_{vec}$ is a column vector with $n$th element given by $P^{vec}_n=P_n$, and $M$ is a tridiagonal matrix with 
\begin{align}
& M_{n,n} = i\Omega_n P_n+\epsilon^2\left(A_n + K_n +  G^{(1)}_{n-1} - G^{(2)}_{n-1}\right),\nonumber \\
&\forall~1\leq n \leq n_{max}-1, \nonumber \\
& M_{n,n+1}=\epsilon^2 A_{n+1},~\forall~n, \nonumber \\
&M_{n-1,n}=\epsilon^2 K_{n-1},~\forall~2\leq n \leq n_{max}.
\end{align}
We can diagonalize the matrix $M$ as
\begin{align}
M = V^{-1} \Lambda V,
\end{align}
where $\Lambda=diag\{\lambda_n\}$ is the diagonal matrix containing the eigenvalues of $M$, and the columns of $V$ are the right eigenvectors of $M$. Formally solving Eq.\eqref{Pn_matrix_equation}, obtaining $g^{(1)}(\tau)$ (Eq.\eqref{Pn_def}) and calculating $S(\omega)$ (Eq.\eqref{spectrum_def}), we get the following expression for the spectrum
\begin{align}
\label{spectrum_expression}
S(\omega) = \sum_{j,\ell,n=1}^{n_{max}} V_{j \ell} (V^{-1})_{\ell n} n \rho_n {\rm Re}\left[ \frac{1}{\lambda_\ell - i\omega} \right].
\end{align}
Using the mathematical relation
\begin{align}
\int_{-\infty}^{\infty} \frac{d\omega}{\pi}{\rm Re}\left[ \frac{1}{i\omega_0+\kappa- i\omega} \right]=1,
\end{align}
for $\omega_0,\kappa>0$, we can confirm that Eq.\eqref{spectrum_expression} satisfies the sum rule in Eq.\eqref{sum_rule}, provided the cut-offs $\omega_{min}$ and $\omega_{max}$ are sufficiently far from relevant system frequencies. We have numerically checked in our calculations that restricting the frequency integral within the cut-offs satisfies the sum rule to a good accuracy.

In calculating the spectrum, $S(\omega)$, we need to integrate over $\tau$. However, the regression formula is valid only for $\tau\gg \tau_M$. This brings into question the accuracy of the  spectrum, calculated according to \eqref{spectrum_expression}. However, the inaccuracy at short times only affect large frequency scales, while the pronounced peaks in the spectra, should be governed by the large $\tau$ regime of $g^{(1)}(\tau)$. Hence we expect the position and shape of peaks of $S(\omega)$ obtained using quantum regression formula to be reasonably accurate for our chosen parameter regime.

\bibliography{ref}

\begin{thebibliography}{92}%
\makeatletter
\providecommand \@ifxundefined [1]{%
 \@ifx{#1\undefined}
}%
\providecommand \@ifnum [1]{%
 \ifnum #1\expandafter \@firstoftwo
 \else \expandafter \@secondoftwo
 \fi
}%
\providecommand \@ifx [1]{%
 \ifx #1\expandafter \@firstoftwo
 \else \expandafter \@secondoftwo
 \fi
}%
\providecommand \natexlab [1]{#1}%
\providecommand \enquote  [1]{``#1''}%
\providecommand \bibnamefont  [1]{#1}%
\providecommand \bibfnamefont [1]{#1}%
\providecommand \citenamefont [1]{#1}%
\providecommand \href@noop [0]{\@secondoftwo}%
\providecommand \href [0]{\begingroup \@sanitize@url \@href}%
\providecommand \@href[1]{\@@startlink{#1}\@@href}%
\providecommand \@@href[1]{\endgroup#1\@@endlink}%
\providecommand \@sanitize@url [0]{\catcode `\\12\catcode `\$12\catcode
  `\&12\catcode `\#12\catcode `\^12\catcode `\_12\catcode `\%12\relax}%
\providecommand \@@startlink[1]{}%
\providecommand \@@endlink[0]{}%
\providecommand \url  [0]{\begingroup\@sanitize@url \@url }%
\providecommand \@url [1]{\endgroup\@href {#1}{\urlprefix }}%
\providecommand \urlprefix  [0]{URL }%
\providecommand \Eprint [0]{\href }%
\providecommand \doibase [0]{http://dx.doi.org/}%
\providecommand \selectlanguage [0]{\@gobble}%
\providecommand \bibinfo  [0]{\@secondoftwo}%
\providecommand \bibfield  [0]{\@secondoftwo}%
\providecommand \translation [1]{[#1]}%
\providecommand \BibitemOpen [0]{}%
\providecommand \bibitemStop [0]{}%
\providecommand \bibitemNoStop [0]{.\EOS\space}%
\providecommand \EOS [0]{\spacefactor3000\relax}%
\providecommand \BibitemShut  [1]{\csname bibitem#1\endcsname}%
\let\auto@bib@innerbib\@empty
\bibitem [{\citenamefont {Devoret}(1997)}]{Devoret_1997}%
  \BibitemOpen
  \bibfield  {author} {\bibinfo {author} {\bibfnamefont {M.~H.}\ \bibnamefont
  {Devoret}},\ }\href
  {http://inis.iaea.org/search/search.aspx?orig_q=RN:29063476} {\emph {\bibinfo
  {title} {Quantum fluctuations in electrical circuits}}}\ (\bibinfo
  {publisher} {Edition de Physique},\ \bibinfo {address} {France},\ \bibinfo
  {year} {1997})\BibitemShut {NoStop}%
\bibitem [{\citenamefont {Clerk}\ \emph {et~al.}(2010)\citenamefont {Clerk},
  \citenamefont {Devoret}, \citenamefont {Girvin}, \citenamefont {Marquardt},\
  and\ \citenamefont {Schoelkopf}}]{Clerk_2010}%
  \BibitemOpen
  \bibfield  {author} {\bibinfo {author} {\bibfnamefont {A.~A.}\ \bibnamefont
  {Clerk}}, \bibinfo {author} {\bibfnamefont {M.~H.}\ \bibnamefont {Devoret}},
  \bibinfo {author} {\bibfnamefont {S.~M.}\ \bibnamefont {Girvin}}, \bibinfo
  {author} {\bibfnamefont {F.}~\bibnamefont {Marquardt}}, \ and\ \bibinfo
  {author} {\bibfnamefont {R.~J.}\ \bibnamefont {Schoelkopf}},\ }\href
  {\doibase 10.1103/RevModPhys.82.1155} {\bibfield  {journal} {\bibinfo
  {journal} {Rev. Mod. Phys.}\ }\textbf {\bibinfo {volume} {82}},\ \bibinfo
  {pages} {1155} (\bibinfo {year} {2010})}\BibitemShut {NoStop}%
\bibitem [{\citenamefont {You}\ \emph {et~al.}(2021)\citenamefont {You},
  \citenamefont {Clerk},\ and\ \citenamefont {Koch}}]{You_2021}%
  \BibitemOpen
  \bibfield  {author} {\bibinfo {author} {\bibfnamefont {X.}~\bibnamefont
  {You}}, \bibinfo {author} {\bibfnamefont {A.~A.}\ \bibnamefont {Clerk}}, \
  and\ \bibinfo {author} {\bibfnamefont {J.}~\bibnamefont {Koch}},\ }\href
  {\doibase 10.1103/PhysRevResearch.3.013045} {\bibfield  {journal} {\bibinfo
  {journal} {Phys. Rev. Res.}\ }\textbf {\bibinfo {volume} {3}},\ \bibinfo
  {pages} {013045} (\bibinfo {year} {2021})}\BibitemShut {NoStop}%
\bibitem [{\citenamefont {Groszkowski}\ \emph {et~al.}(2023)\citenamefont
  {Groszkowski}, \citenamefont {Seif}, \citenamefont {Koch},\ and\
  \citenamefont {Clerk}}]{Groszkowski_2023}%
  \BibitemOpen
  \bibfield  {author} {\bibinfo {author} {\bibfnamefont {P.}~\bibnamefont
  {Groszkowski}}, \bibinfo {author} {\bibfnamefont {A.}~\bibnamefont {Seif}},
  \bibinfo {author} {\bibfnamefont {J.}~\bibnamefont {Koch}}, \ and\ \bibinfo
  {author} {\bibfnamefont {A.~A.}\ \bibnamefont {Clerk}},\ }\href {\doibase
  10.22331/q-2023-04-06-972} {\bibfield  {journal} {\bibinfo  {journal}
  {Quantum}\ }\textbf {\bibinfo {volume} {7}},\ \bibinfo {pages} {972}
  (\bibinfo {year} {2023})}\BibitemShut {NoStop}%
\bibitem [{\citenamefont {Paladino}\ \emph {et~al.}(2014)\citenamefont
  {Paladino}, \citenamefont {Galperin}, \citenamefont {Falci},\ and\
  \citenamefont {Altshuler}}]{Paladino_2014}%
  \BibitemOpen
  \bibfield  {author} {\bibinfo {author} {\bibfnamefont {E.}~\bibnamefont
  {Paladino}}, \bibinfo {author} {\bibfnamefont {Y.~M.}\ \bibnamefont
  {Galperin}}, \bibinfo {author} {\bibfnamefont {G.}~\bibnamefont {Falci}}, \
  and\ \bibinfo {author} {\bibfnamefont {B.~L.}\ \bibnamefont {Altshuler}},\
  }\href {\doibase 10.1103/RevModPhys.86.361} {\bibfield  {journal} {\bibinfo
  {journal} {Rev. Mod. Phys.}\ }\textbf {\bibinfo {volume} {86}},\ \bibinfo
  {pages} {361} (\bibinfo {year} {2014})}\BibitemShut {NoStop}%
\bibitem [{\citenamefont {Braum\"uller}\ \emph {et~al.}(2020)\citenamefont
  {Braum\"uller}, \citenamefont {Ding}, \citenamefont {Veps\"al\"ainen},
  \citenamefont {Sung}, \citenamefont {Kjaergaard}, \citenamefont {Menke},
  \citenamefont {Winik}, \citenamefont {Kim}, \citenamefont {Niedzielski},
  \citenamefont {Melville}, \citenamefont {Yoder}, \citenamefont
  {Hirjibehedin}, \citenamefont {Orlando}, \citenamefont {Gustavsson},\ and\
  \citenamefont {Oliver}}]{Braumuller_2020}%
  \BibitemOpen
  \bibfield  {author} {\bibinfo {author} {\bibfnamefont {J.}~\bibnamefont
  {Braum\"uller}}, \bibinfo {author} {\bibfnamefont {L.}~\bibnamefont {Ding}},
  \bibinfo {author} {\bibfnamefont {A.~P.}\ \bibnamefont {Veps\"al\"ainen}},
  \bibinfo {author} {\bibfnamefont {Y.}~\bibnamefont {Sung}}, \bibinfo {author}
  {\bibfnamefont {M.}~\bibnamefont {Kjaergaard}}, \bibinfo {author}
  {\bibfnamefont {T.}~\bibnamefont {Menke}}, \bibinfo {author} {\bibfnamefont
  {R.}~\bibnamefont {Winik}}, \bibinfo {author} {\bibfnamefont
  {D.}~\bibnamefont {Kim}}, \bibinfo {author} {\bibfnamefont {B.~M.}\
  \bibnamefont {Niedzielski}}, \bibinfo {author} {\bibfnamefont
  {A.}~\bibnamefont {Melville}}, \bibinfo {author} {\bibfnamefont {J.~L.}\
  \bibnamefont {Yoder}}, \bibinfo {author} {\bibfnamefont {C.~F.}\ \bibnamefont
  {Hirjibehedin}}, \bibinfo {author} {\bibfnamefont {T.~P.}\ \bibnamefont
  {Orlando}}, \bibinfo {author} {\bibfnamefont {S.}~\bibnamefont {Gustavsson}},
  \ and\ \bibinfo {author} {\bibfnamefont {W.~D.}\ \bibnamefont {Oliver}},\
  }\href {\doibase 10.1103/PhysRevApplied.13.054079} {\bibfield  {journal}
  {\bibinfo  {journal} {Phys. Rev. Appl.}\ }\textbf {\bibinfo {volume} {13}},\
  \bibinfo {pages} {054079} (\bibinfo {year} {2020})}\BibitemShut {NoStop}%
\bibitem [{\citenamefont {Kuhlmann}\ \emph {et~al.}(2013)\citenamefont
  {Kuhlmann}, \citenamefont {Houel}, \citenamefont {Ludwig}, \citenamefont
  {Greuter}, \citenamefont {Reuter}, \citenamefont {Wieck}, \citenamefont
  {Poggio},\ and\ \citenamefont {Warburton}}]{Kuhlmann_2013}%
  \BibitemOpen
  \bibfield  {author} {\bibinfo {author} {\bibfnamefont {A.~V.}\ \bibnamefont
  {Kuhlmann}}, \bibinfo {author} {\bibfnamefont {J.}~\bibnamefont {Houel}},
  \bibinfo {author} {\bibfnamefont {A.}~\bibnamefont {Ludwig}}, \bibinfo
  {author} {\bibfnamefont {L.}~\bibnamefont {Greuter}}, \bibinfo {author}
  {\bibfnamefont {D.}~\bibnamefont {Reuter}}, \bibinfo {author} {\bibfnamefont
  {A.~D.}\ \bibnamefont {Wieck}}, \bibinfo {author} {\bibfnamefont
  {M.}~\bibnamefont {Poggio}}, \ and\ \bibinfo {author} {\bibfnamefont {R.~J.}\
  \bibnamefont {Warburton}},\ }\href {\doibase 10.1038/nphys2688} {\bibfield
  {journal} {\bibinfo  {journal} {Nature Physics}\ }\textbf {\bibinfo {volume}
  {9}},\ \bibinfo {pages} {570} (\bibinfo {year} {2013})}\BibitemShut {NoStop}%
\bibitem [{\citenamefont {Yoshihara}\ \emph {et~al.}(2006)\citenamefont
  {Yoshihara}, \citenamefont {Harrabi}, \citenamefont {Niskanen}, \citenamefont
  {Nakamura},\ and\ \citenamefont {Tsai}}]{Yoshihara_2006}%
  \BibitemOpen
  \bibfield  {author} {\bibinfo {author} {\bibfnamefont {F.}~\bibnamefont
  {Yoshihara}}, \bibinfo {author} {\bibfnamefont {K.}~\bibnamefont {Harrabi}},
  \bibinfo {author} {\bibfnamefont {A.~O.}\ \bibnamefont {Niskanen}}, \bibinfo
  {author} {\bibfnamefont {Y.}~\bibnamefont {Nakamura}}, \ and\ \bibinfo
  {author} {\bibfnamefont {J.~S.}\ \bibnamefont {Tsai}},\ }\href {\doibase
  10.1103/PhysRevLett.97.167001} {\bibfield  {journal} {\bibinfo  {journal}
  {Phys. Rev. Lett.}\ }\textbf {\bibinfo {volume} {97}},\ \bibinfo {pages}
  {167001} (\bibinfo {year} {2006})}\BibitemShut {NoStop}%
\bibitem [{\citenamefont {Quintana}\ \emph {et~al.}(2017)\citenamefont
  {Quintana}, \citenamefont {Chen}, \citenamefont {Sank}, \citenamefont
  {Petukhov}, \citenamefont {White}, \citenamefont {Kafri}, \citenamefont
  {Chiaro}, \citenamefont {Megrant}, \citenamefont {Barends}, \citenamefont
  {Campbell}, \citenamefont {Chen}, \citenamefont {Dunsworth}, \citenamefont
  {Fowler}, \citenamefont {Graff}, \citenamefont {Jeffrey}, \citenamefont
  {Kelly}, \citenamefont {Lucero}, \citenamefont {Mutus}, \citenamefont
  {Neeley}, \citenamefont {Neill}, \citenamefont {O'Malley}, \citenamefont
  {Roushan}, \citenamefont {Shabani}, \citenamefont {Smelyanskiy},
  \citenamefont {Vainsencher}, \citenamefont {Wenner}, \citenamefont {Neven},\
  and\ \citenamefont {Martinis}}]{Quintana_2017}%
  \BibitemOpen
  \bibfield  {author} {\bibinfo {author} {\bibfnamefont {C.~M.}\ \bibnamefont
  {Quintana}}, \bibinfo {author} {\bibfnamefont {Y.}~\bibnamefont {Chen}},
  \bibinfo {author} {\bibfnamefont {D.}~\bibnamefont {Sank}}, \bibinfo {author}
  {\bibfnamefont {A.~G.}\ \bibnamefont {Petukhov}}, \bibinfo {author}
  {\bibfnamefont {T.~C.}\ \bibnamefont {White}}, \bibinfo {author}
  {\bibfnamefont {D.}~\bibnamefont {Kafri}}, \bibinfo {author} {\bibfnamefont
  {B.}~\bibnamefont {Chiaro}}, \bibinfo {author} {\bibfnamefont
  {A.}~\bibnamefont {Megrant}}, \bibinfo {author} {\bibfnamefont
  {R.}~\bibnamefont {Barends}}, \bibinfo {author} {\bibfnamefont
  {B.}~\bibnamefont {Campbell}}, \bibinfo {author} {\bibfnamefont
  {Z.}~\bibnamefont {Chen}}, \bibinfo {author} {\bibfnamefont {A.}~\bibnamefont
  {Dunsworth}}, \bibinfo {author} {\bibfnamefont {A.~G.}\ \bibnamefont
  {Fowler}}, \bibinfo {author} {\bibfnamefont {R.}~\bibnamefont {Graff}},
  \bibinfo {author} {\bibfnamefont {E.}~\bibnamefont {Jeffrey}}, \bibinfo
  {author} {\bibfnamefont {J.}~\bibnamefont {Kelly}}, \bibinfo {author}
  {\bibfnamefont {E.}~\bibnamefont {Lucero}}, \bibinfo {author} {\bibfnamefont
  {J.~Y.}\ \bibnamefont {Mutus}}, \bibinfo {author} {\bibfnamefont
  {M.}~\bibnamefont {Neeley}}, \bibinfo {author} {\bibfnamefont
  {C.}~\bibnamefont {Neill}}, \bibinfo {author} {\bibfnamefont {P.~J.~J.}\
  \bibnamefont {O'Malley}}, \bibinfo {author} {\bibfnamefont {P.}~\bibnamefont
  {Roushan}}, \bibinfo {author} {\bibfnamefont {A.}~\bibnamefont {Shabani}},
  \bibinfo {author} {\bibfnamefont {V.~N.}\ \bibnamefont {Smelyanskiy}},
  \bibinfo {author} {\bibfnamefont {A.}~\bibnamefont {Vainsencher}}, \bibinfo
  {author} {\bibfnamefont {J.}~\bibnamefont {Wenner}}, \bibinfo {author}
  {\bibfnamefont {H.}~\bibnamefont {Neven}}, \ and\ \bibinfo {author}
  {\bibfnamefont {J.~M.}\ \bibnamefont {Martinis}},\ }\href {\doibase
  10.1103/PhysRevLett.118.057702} {\bibfield  {journal} {\bibinfo  {journal}
  {Phys. Rev. Lett.}\ }\textbf {\bibinfo {volume} {118}},\ \bibinfo {pages}
  {057702} (\bibinfo {year} {2017})}\BibitemShut {NoStop}%
\bibitem [{\citenamefont {Norris}\ \emph {et~al.}(2016)\citenamefont {Norris},
  \citenamefont {Paz-Silva},\ and\ \citenamefont {Viola}}]{Norris_2016}%
  \BibitemOpen
  \bibfield  {author} {\bibinfo {author} {\bibfnamefont {L.~M.}\ \bibnamefont
  {Norris}}, \bibinfo {author} {\bibfnamefont {G.~A.}\ \bibnamefont
  {Paz-Silva}}, \ and\ \bibinfo {author} {\bibfnamefont {L.}~\bibnamefont
  {Viola}},\ }\href {\doibase 10.1103/PhysRevLett.116.150503} {\bibfield
  {journal} {\bibinfo  {journal} {Phys. Rev. Lett.}\ }\textbf {\bibinfo
  {volume} {116}},\ \bibinfo {pages} {150503} (\bibinfo {year}
  {2016})}\BibitemShut {NoStop}%
\bibitem [{\citenamefont {O'Malley}\ \emph {et~al.}(2015)\citenamefont
  {O'Malley}, \citenamefont {Kelly}, \citenamefont {Barends}, \citenamefont
  {Campbell}, \citenamefont {Chen}, \citenamefont {Chen}, \citenamefont
  {Chiaro}, \citenamefont {Dunsworth}, \citenamefont {Fowler}, \citenamefont
  {Hoi}, \citenamefont {Jeffrey}, \citenamefont {Megrant}, \citenamefont
  {Mutus}, \citenamefont {Neill}, \citenamefont {Quintana}, \citenamefont
  {Roushan}, \citenamefont {Sank}, \citenamefont {Vainsencher}, \citenamefont
  {Wenner}, \citenamefont {White}, \citenamefont {Korotkov}, \citenamefont
  {Cleland},\ and\ \citenamefont {Martinis}}]{OMalley_2015}%
  \BibitemOpen
  \bibfield  {author} {\bibinfo {author} {\bibfnamefont {P.~J.~J.}\
  \bibnamefont {O'Malley}}, \bibinfo {author} {\bibfnamefont {J.}~\bibnamefont
  {Kelly}}, \bibinfo {author} {\bibfnamefont {R.}~\bibnamefont {Barends}},
  \bibinfo {author} {\bibfnamefont {B.}~\bibnamefont {Campbell}}, \bibinfo
  {author} {\bibfnamefont {Y.}~\bibnamefont {Chen}}, \bibinfo {author}
  {\bibfnamefont {Z.}~\bibnamefont {Chen}}, \bibinfo {author} {\bibfnamefont
  {B.}~\bibnamefont {Chiaro}}, \bibinfo {author} {\bibfnamefont
  {A.}~\bibnamefont {Dunsworth}}, \bibinfo {author} {\bibfnamefont {A.~G.}\
  \bibnamefont {Fowler}}, \bibinfo {author} {\bibfnamefont {I.-C.}\
  \bibnamefont {Hoi}}, \bibinfo {author} {\bibfnamefont {E.}~\bibnamefont
  {Jeffrey}}, \bibinfo {author} {\bibfnamefont {A.}~\bibnamefont {Megrant}},
  \bibinfo {author} {\bibfnamefont {J.}~\bibnamefont {Mutus}}, \bibinfo
  {author} {\bibfnamefont {C.}~\bibnamefont {Neill}}, \bibinfo {author}
  {\bibfnamefont {C.}~\bibnamefont {Quintana}}, \bibinfo {author}
  {\bibfnamefont {P.}~\bibnamefont {Roushan}}, \bibinfo {author} {\bibfnamefont
  {D.}~\bibnamefont {Sank}}, \bibinfo {author} {\bibfnamefont {A.}~\bibnamefont
  {Vainsencher}}, \bibinfo {author} {\bibfnamefont {J.}~\bibnamefont {Wenner}},
  \bibinfo {author} {\bibfnamefont {T.~C.}\ \bibnamefont {White}}, \bibinfo
  {author} {\bibfnamefont {A.~N.}\ \bibnamefont {Korotkov}}, \bibinfo {author}
  {\bibfnamefont {A.~N.}\ \bibnamefont {Cleland}}, \ and\ \bibinfo {author}
  {\bibfnamefont {J.~M.}\ \bibnamefont {Martinis}},\ }\href {\doibase
  10.1103/PhysRevApplied.3.044009} {\bibfield  {journal} {\bibinfo  {journal}
  {Phys. Rev. Appl.}\ }\textbf {\bibinfo {volume} {3}},\ \bibinfo {pages}
  {044009} (\bibinfo {year} {2015})}\BibitemShut {NoStop}%
\bibitem [{\citenamefont {\'Alvarez}\ and\ \citenamefont
  {Suter}(2011)}]{Alvarez_2011}%
  \BibitemOpen
  \bibfield  {author} {\bibinfo {author} {\bibfnamefont {G.~A.}\ \bibnamefont
  {\'Alvarez}}\ and\ \bibinfo {author} {\bibfnamefont {D.}~\bibnamefont
  {Suter}},\ }\href {\doibase 10.1103/PhysRevLett.107.230501} {\bibfield
  {journal} {\bibinfo  {journal} {Phys. Rev. Lett.}\ }\textbf {\bibinfo
  {volume} {107}},\ \bibinfo {pages} {230501} (\bibinfo {year}
  {2011})}\BibitemShut {NoStop}%
\bibitem [{\citenamefont {Bylander}\ \emph {et~al.}(2011)\citenamefont
  {Bylander}, \citenamefont {Gustavsson}, \citenamefont {Yan}, \citenamefont
  {Yoshihara}, \citenamefont {Harrabi}, \citenamefont {Fitch}, \citenamefont
  {Cory}, \citenamefont {Nakamura}, \citenamefont {Tsai},\ and\ \citenamefont
  {Oliver}}]{Bylander_2011}%
  \BibitemOpen
  \bibfield  {author} {\bibinfo {author} {\bibfnamefont {J.}~\bibnamefont
  {Bylander}}, \bibinfo {author} {\bibfnamefont {S.}~\bibnamefont
  {Gustavsson}}, \bibinfo {author} {\bibfnamefont {F.}~\bibnamefont {Yan}},
  \bibinfo {author} {\bibfnamefont {F.}~\bibnamefont {Yoshihara}}, \bibinfo
  {author} {\bibfnamefont {K.}~\bibnamefont {Harrabi}}, \bibinfo {author}
  {\bibfnamefont {G.}~\bibnamefont {Fitch}}, \bibinfo {author} {\bibfnamefont
  {D.~G.}\ \bibnamefont {Cory}}, \bibinfo {author} {\bibfnamefont
  {Y.}~\bibnamefont {Nakamura}}, \bibinfo {author} {\bibfnamefont {J.-S.}\
  \bibnamefont {Tsai}}, \ and\ \bibinfo {author} {\bibfnamefont {W.~D.}\
  \bibnamefont {Oliver}},\ }\href {\doibase 10.1038/nphys1994} {\bibfield
  {journal} {\bibinfo  {journal} {Nature Physics}\ }\textbf {\bibinfo {volume}
  {7}},\ \bibinfo {pages} {565} (\bibinfo {year} {2011})}\BibitemShut {NoStop}%
\bibitem [{\citenamefont {Shor}(1995)}]{Shor_1995}%
  \BibitemOpen
  \bibfield  {author} {\bibinfo {author} {\bibfnamefont {P.~W.}\ \bibnamefont
  {Shor}},\ }\href {\doibase 10.1103/PhysRevA.52.R2493} {\bibfield  {journal}
  {\bibinfo  {journal} {Phys. Rev. A}\ }\textbf {\bibinfo {volume} {52}},\
  \bibinfo {pages} {R2493} (\bibinfo {year} {1995})}\BibitemShut {NoStop}%
\bibitem [{\citenamefont {Viola}\ \emph {et~al.}(1999)\citenamefont {Viola},
  \citenamefont {Knill},\ and\ \citenamefont {Lloyd}}]{Viola_1999}%
  \BibitemOpen
  \bibfield  {author} {\bibinfo {author} {\bibfnamefont {L.}~\bibnamefont
  {Viola}}, \bibinfo {author} {\bibfnamefont {E.}~\bibnamefont {Knill}}, \ and\
  \bibinfo {author} {\bibfnamefont {S.}~\bibnamefont {Lloyd}},\ }\href
  {\doibase 10.1103/PhysRevLett.82.2417} {\bibfield  {journal} {\bibinfo
  {journal} {Phys. Rev. Lett.}\ }\textbf {\bibinfo {volume} {82}},\ \bibinfo
  {pages} {2417} (\bibinfo {year} {1999})}\BibitemShut {NoStop}%
\bibitem [{\citenamefont {Rabitz}\ \emph {et~al.}(2000)\citenamefont {Rabitz},
  \citenamefont {de~Vivie-Riedle}, \citenamefont {Motzkus},\ and\ \citenamefont
  {Kompa}}]{Rabitz_2000}%
  \BibitemOpen
  \bibfield  {author} {\bibinfo {author} {\bibfnamefont {H.}~\bibnamefont
  {Rabitz}}, \bibinfo {author} {\bibfnamefont {R.}~\bibnamefont
  {de~Vivie-Riedle}}, \bibinfo {author} {\bibfnamefont {M.}~\bibnamefont
  {Motzkus}}, \ and\ \bibinfo {author} {\bibfnamefont {K.}~\bibnamefont
  {Kompa}},\ }\href {\doibase 10.1126/science.288.5467.824} {\bibfield
  {journal} {\bibinfo  {journal} {Science}\ }\textbf {\bibinfo {volume}
  {288}},\ \bibinfo {pages} {824} (\bibinfo {year} {2000})}\BibitemShut
  {NoStop}%
\bibitem [{\citenamefont {Faoro}\ and\ \citenamefont
  {Viola}(2004)}]{Faoro_2004}%
  \BibitemOpen
  \bibfield  {author} {\bibinfo {author} {\bibfnamefont {L.}~\bibnamefont
  {Faoro}}\ and\ \bibinfo {author} {\bibfnamefont {L.}~\bibnamefont {Viola}},\
  }\href {\doibase 10.1103/PhysRevLett.92.117905} {\bibfield  {journal}
  {\bibinfo  {journal} {Phys. Rev. Lett.}\ }\textbf {\bibinfo {volume} {92}},\
  \bibinfo {pages} {117905} (\bibinfo {year} {2004})}\BibitemShut {NoStop}%
\bibitem [{\citenamefont {Yang}\ \emph {et~al.}(2019)\citenamefont {Yang},
  \citenamefont {Coppersmith},\ and\ \citenamefont {Friesen}}]{Yang_2019}%
  \BibitemOpen
  \bibfield  {author} {\bibinfo {author} {\bibfnamefont {Y.-C.}\ \bibnamefont
  {Yang}}, \bibinfo {author} {\bibfnamefont {S.~N.}\ \bibnamefont
  {Coppersmith}}, \ and\ \bibinfo {author} {\bibfnamefont {M.}~\bibnamefont
  {Friesen}},\ }\href {\doibase 10.1038/s41534-019-0127-1} {\bibfield
  {journal} {\bibinfo  {journal} {npj Quantum Information}\ }\textbf {\bibinfo
  {volume} {5}},\ \bibinfo {pages} {12} (\bibinfo {year} {2019})}\BibitemShut
  {NoStop}%
\bibitem [{\citenamefont {Souza}\ \emph {et~al.}(2012)\citenamefont {Souza},
  \citenamefont {Álvarez},\ and\ \citenamefont {Suter}}]{Sourza_2012}%
  \BibitemOpen
  \bibfield  {author} {\bibinfo {author} {\bibfnamefont {A.~M.}\ \bibnamefont
  {Souza}}, \bibinfo {author} {\bibfnamefont {G.~A.}\ \bibnamefont {Álvarez}},
  \ and\ \bibinfo {author} {\bibfnamefont {D.}~\bibnamefont {Suter}},\ }\href
  {\doibase 10.1098/rsta.2011.0355} {\bibfield  {journal} {\bibinfo  {journal}
  {Philosophical Transactions of the Royal Society A: Mathematical, Physical
  and Engineering Sciences}\ }\textbf {\bibinfo {volume} {370}},\ \bibinfo
  {pages} {4748} (\bibinfo {year} {2012})}\BibitemShut {NoStop}%
\bibitem [{\citenamefont {Malinowski}\ \emph {et~al.}(2017)\citenamefont
  {Malinowski}, \citenamefont {Martins}, \citenamefont {Nissen}, \citenamefont
  {Barnes}, \citenamefont {Cywi{\'{n}}ski}, \citenamefont {Rudner},
  \citenamefont {Fallahi}, \citenamefont {Gardner}, \citenamefont {Manfra},
  \citenamefont {Marcus},\ and\ \citenamefont {Kuemmeth}}]{Malinowski_2017}%
  \BibitemOpen
  \bibfield  {author} {\bibinfo {author} {\bibfnamefont {F.~K.}\ \bibnamefont
  {Malinowski}}, \bibinfo {author} {\bibfnamefont {F.}~\bibnamefont {Martins}},
  \bibinfo {author} {\bibfnamefont {P.~D.}\ \bibnamefont {Nissen}}, \bibinfo
  {author} {\bibfnamefont {E.}~\bibnamefont {Barnes}}, \bibinfo {author}
  {\bibfnamefont {{\L}.}~\bibnamefont {Cywi{\'{n}}ski}}, \bibinfo {author}
  {\bibfnamefont {M.~S.}\ \bibnamefont {Rudner}}, \bibinfo {author}
  {\bibfnamefont {S.}~\bibnamefont {Fallahi}}, \bibinfo {author} {\bibfnamefont
  {G.~C.}\ \bibnamefont {Gardner}}, \bibinfo {author} {\bibfnamefont {M.~J.}\
  \bibnamefont {Manfra}}, \bibinfo {author} {\bibfnamefont {C.~M.}\
  \bibnamefont {Marcus}}, \ and\ \bibinfo {author} {\bibfnamefont
  {F.}~\bibnamefont {Kuemmeth}},\ }\href {\doibase 10.1038/nnano.2016.170}
  {\bibfield  {journal} {\bibinfo  {journal} {Nature Nanotechnology}\ }\textbf
  {\bibinfo {volume} {12}},\ \bibinfo {pages} {16} (\bibinfo {year}
  {2017})}\BibitemShut {NoStop}%
\bibitem [{\citenamefont {Baum}\ \emph {et~al.}(2021)\citenamefont {Baum},
  \citenamefont {Amico}, \citenamefont {Howell}, \citenamefont {Hush},
  \citenamefont {Liuzzi}, \citenamefont {Mundada}, \citenamefont {Merkh},
  \citenamefont {Carvalho},\ and\ \citenamefont {Biercuk}}]{Baum_2021}%
  \BibitemOpen
  \bibfield  {author} {\bibinfo {author} {\bibfnamefont {Y.}~\bibnamefont
  {Baum}}, \bibinfo {author} {\bibfnamefont {M.}~\bibnamefont {Amico}},
  \bibinfo {author} {\bibfnamefont {S.}~\bibnamefont {Howell}}, \bibinfo
  {author} {\bibfnamefont {M.}~\bibnamefont {Hush}}, \bibinfo {author}
  {\bibfnamefont {M.}~\bibnamefont {Liuzzi}}, \bibinfo {author} {\bibfnamefont
  {P.}~\bibnamefont {Mundada}}, \bibinfo {author} {\bibfnamefont
  {T.}~\bibnamefont {Merkh}}, \bibinfo {author} {\bibfnamefont {A.~R.}\
  \bibnamefont {Carvalho}}, \ and\ \bibinfo {author} {\bibfnamefont {M.~J.}\
  \bibnamefont {Biercuk}},\ }\href {\doibase 10.1103/PRXQuantum.2.040324}
  {\bibfield  {journal} {\bibinfo  {journal} {PRX Quantum}\ }\textbf {\bibinfo
  {volume} {2}},\ \bibinfo {pages} {040324} (\bibinfo {year}
  {2021})}\BibitemShut {NoStop}%
\bibitem [{\citenamefont {Ezzell}\ \emph {et~al.}(2023)\citenamefont {Ezzell},
  \citenamefont {Pokharel}, \citenamefont {Tewala}, \citenamefont {Quiroz},\
  and\ \citenamefont {Lidar}}]{Ezzell_2023}%
  \BibitemOpen
  \bibfield  {author} {\bibinfo {author} {\bibfnamefont {N.}~\bibnamefont
  {Ezzell}}, \bibinfo {author} {\bibfnamefont {B.}~\bibnamefont {Pokharel}},
  \bibinfo {author} {\bibfnamefont {L.}~\bibnamefont {Tewala}}, \bibinfo
  {author} {\bibfnamefont {G.}~\bibnamefont {Quiroz}}, \ and\ \bibinfo {author}
  {\bibfnamefont {D.~A.}\ \bibnamefont {Lidar}},\ }\href@noop {} {\  (\bibinfo
  {year} {2023})},\ \Eprint {http://arxiv.org/abs/2207.03670} {arXiv:2207.03670
  [quant-ph]} \BibitemShut {NoStop}%
\bibitem [{\citenamefont {Purkayastha}\ \emph {et~al.}(2020)\citenamefont
  {Purkayastha}, \citenamefont {Guarnieri}, \citenamefont {Mitchison},
  \citenamefont {Filip},\ and\ \citenamefont {Goold}}]{Purkayastha_2020}%
  \BibitemOpen
  \bibfield  {author} {\bibinfo {author} {\bibfnamefont {A.}~\bibnamefont
  {Purkayastha}}, \bibinfo {author} {\bibfnamefont {G.}~\bibnamefont
  {Guarnieri}}, \bibinfo {author} {\bibfnamefont {M.~T.}\ \bibnamefont
  {Mitchison}}, \bibinfo {author} {\bibfnamefont {R.}~\bibnamefont {Filip}}, \
  and\ \bibinfo {author} {\bibfnamefont {J.}~\bibnamefont {Goold}},\ }\href
  {\doibase 10.1038/s41534-020-0256-6} {\bibfield  {journal} {\bibinfo
  {journal} {npj Quantum Information}\ }\textbf {\bibinfo {volume} {6}},\
  \bibinfo {pages} {27} (\bibinfo {year} {2020})}\BibitemShut {NoStop}%
\bibitem [{\citenamefont {Guarnieri}\ \emph {et~al.}(2018)\citenamefont
  {Guarnieri}, \citenamefont {Kol\'a\ifmmode~\check{r}\else \v{r}\fi{}},\ and\
  \citenamefont {Filip}}]{Guarnieri_2018}%
  \BibitemOpen
  \bibfield  {author} {\bibinfo {author} {\bibfnamefont {G.}~\bibnamefont
  {Guarnieri}}, \bibinfo {author} {\bibfnamefont {M.}~\bibnamefont
  {Kol\'a\ifmmode~\check{r}\else \v{r}\fi{}}}, \ and\ \bibinfo {author}
  {\bibfnamefont {R.}~\bibnamefont {Filip}},\ }\href {\doibase
  10.1103/PhysRevLett.121.070401} {\bibfield  {journal} {\bibinfo  {journal}
  {Phys. Rev. Lett.}\ }\textbf {\bibinfo {volume} {121}},\ \bibinfo {pages}
  {070401} (\bibinfo {year} {2018})}\BibitemShut {NoStop}%
\bibitem [{\citenamefont {Ghosh}\ \emph {et~al.}(2021)\citenamefont {Ghosh},
  \citenamefont {Chanda}, \citenamefont {Mal},\ and\ \citenamefont
  {Sen(De)}}]{Ghosh_2021}%
  \BibitemOpen
  \bibfield  {author} {\bibinfo {author} {\bibfnamefont {S.}~\bibnamefont
  {Ghosh}}, \bibinfo {author} {\bibfnamefont {T.}~\bibnamefont {Chanda}},
  \bibinfo {author} {\bibfnamefont {S.}~\bibnamefont {Mal}}, \ and\ \bibinfo
  {author} {\bibfnamefont {A.}~\bibnamefont {Sen(De)}},\ }\href {\doibase
  10.1103/PhysRevA.104.032207} {\bibfield  {journal} {\bibinfo  {journal}
  {Phys. Rev. A}\ }\textbf {\bibinfo {volume} {104}},\ \bibinfo {pages}
  {032207} (\bibinfo {year} {2021})}\BibitemShut {NoStop}%
\bibitem [{\citenamefont {Cao}\ and\ \citenamefont {Wang}(2021)}]{Cao_2021}%
  \BibitemOpen
  \bibfield  {author} {\bibinfo {author} {\bibfnamefont {C.}~\bibnamefont
  {Cao}}\ and\ \bibinfo {author} {\bibfnamefont {X.}~\bibnamefont {Wang}},\
  }\href {\doibase 10.1103/PhysRevApplied.15.054012} {\bibfield  {journal}
  {\bibinfo  {journal} {Phys. Rev. Appl.}\ }\textbf {\bibinfo {volume} {15}},\
  \bibinfo {pages} {054012} (\bibinfo {year} {2021})}\BibitemShut {NoStop}%
\bibitem [{\citenamefont {Aamir}\ \emph {et~al.}(2023)\citenamefont {Aamir},
  \citenamefont {Suria}, \citenamefont {Guzmán}, \citenamefont
  {Castillo-Moreno}, \citenamefont {Epstein}, \citenamefont {Halpern},\ and\
  \citenamefont {Gasparinetti}}]{Aamir_2023}%
  \BibitemOpen
  \bibfield  {author} {\bibinfo {author} {\bibfnamefont {M.~A.}\ \bibnamefont
  {Aamir}}, \bibinfo {author} {\bibfnamefont {P.~J.}\ \bibnamefont {Suria}},
  \bibinfo {author} {\bibfnamefont {J.~A.~M.}\ \bibnamefont {Guzmán}},
  \bibinfo {author} {\bibfnamefont {C.}~\bibnamefont {Castillo-Moreno}},
  \bibinfo {author} {\bibfnamefont {J.~M.}\ \bibnamefont {Epstein}}, \bibinfo
  {author} {\bibfnamefont {N.~Y.}\ \bibnamefont {Halpern}}, \ and\ \bibinfo
  {author} {\bibfnamefont {S.}~\bibnamefont {Gasparinetti}},\ }\href@noop {} {\
   (\bibinfo {year} {2023})},\ \Eprint {http://arxiv.org/abs/2305.16710}
  {arXiv:2305.16710 [quant-ph]} \BibitemShut {NoStop}%
\bibitem [{\citenamefont {Wang}\ \emph {et~al.}(2023)\citenamefont {Wang},
  \citenamefont {Banniard}, \citenamefont {Børkje}, \citenamefont {Massel},
  \citenamefont {de~Lépinay},\ and\ \citenamefont {Sillanpää}}]{Wang_2023}%
  \BibitemOpen
  \bibfield  {author} {\bibinfo {author} {\bibfnamefont {C.}~\bibnamefont
  {Wang}}, \bibinfo {author} {\bibfnamefont {L.}~\bibnamefont {Banniard}},
  \bibinfo {author} {\bibfnamefont {K.}~\bibnamefont {Børkje}}, \bibinfo
  {author} {\bibfnamefont {F.}~\bibnamefont {Massel}}, \bibinfo {author}
  {\bibfnamefont {L.~M.}\ \bibnamefont {de~Lépinay}}, \ and\ \bibinfo {author}
  {\bibfnamefont {M.~A.}\ \bibnamefont {Sillanpää}},\ }\href@noop {} {\
  (\bibinfo {year} {2023})},\ \Eprint {http://arxiv.org/abs/2306.15746}
  {arXiv:2306.15746 [quant-ph]} \BibitemShut {NoStop}%
\bibitem [{\citenamefont {Safavi-Naini}\ \emph {et~al.}(2011)\citenamefont
  {Safavi-Naini}, \citenamefont {Rabl}, \citenamefont {Weck},\ and\
  \citenamefont {Sadeghpour}}]{Safavi-Naini_2011}%
  \BibitemOpen
  \bibfield  {author} {\bibinfo {author} {\bibfnamefont {A.}~\bibnamefont
  {Safavi-Naini}}, \bibinfo {author} {\bibfnamefont {P.}~\bibnamefont {Rabl}},
  \bibinfo {author} {\bibfnamefont {P.~F.}\ \bibnamefont {Weck}}, \ and\
  \bibinfo {author} {\bibfnamefont {H.~R.}\ \bibnamefont {Sadeghpour}},\ }\href
  {\doibase 10.1103/PhysRevA.84.023412} {\bibfield  {journal} {\bibinfo
  {journal} {Phys. Rev. A}\ }\textbf {\bibinfo {volume} {84}},\ \bibinfo
  {pages} {023412} (\bibinfo {year} {2011})}\BibitemShut {NoStop}%
\bibitem [{\citenamefont {Sedlacek}\ \emph {et~al.}(2018)\citenamefont
  {Sedlacek}, \citenamefont {Stuart}, \citenamefont {Slichter}, \citenamefont
  {Bruzewicz}, \citenamefont {McConnell}, \citenamefont {Sage},\ and\
  \citenamefont {Chiaverini}}]{Sedlacek_2018}%
  \BibitemOpen
  \bibfield  {author} {\bibinfo {author} {\bibfnamefont {J.~A.}\ \bibnamefont
  {Sedlacek}}, \bibinfo {author} {\bibfnamefont {J.}~\bibnamefont {Stuart}},
  \bibinfo {author} {\bibfnamefont {D.~H.}\ \bibnamefont {Slichter}}, \bibinfo
  {author} {\bibfnamefont {C.~D.}\ \bibnamefont {Bruzewicz}}, \bibinfo {author}
  {\bibfnamefont {R.}~\bibnamefont {McConnell}}, \bibinfo {author}
  {\bibfnamefont {J.~M.}\ \bibnamefont {Sage}}, \ and\ \bibinfo {author}
  {\bibfnamefont {J.}~\bibnamefont {Chiaverini}},\ }\href {\doibase
  10.1103/PhysRevA.98.063430} {\bibfield  {journal} {\bibinfo  {journal} {Phys.
  Rev. A}\ }\textbf {\bibinfo {volume} {98}},\ \bibinfo {pages} {063430}
  (\bibinfo {year} {2018})}\BibitemShut {NoStop}%
\bibitem [{\citenamefont {Noel}\ \emph {et~al.}(2019)\citenamefont {Noel},
  \citenamefont {Berlin-Udi}, \citenamefont {Matthiesen}, \citenamefont {Yu},
  \citenamefont {Zhou}, \citenamefont {Lordi},\ and\ \citenamefont
  {H\"affner}}]{Noel_2019}%
  \BibitemOpen
  \bibfield  {author} {\bibinfo {author} {\bibfnamefont {C.}~\bibnamefont
  {Noel}}, \bibinfo {author} {\bibfnamefont {M.}~\bibnamefont {Berlin-Udi}},
  \bibinfo {author} {\bibfnamefont {C.}~\bibnamefont {Matthiesen}}, \bibinfo
  {author} {\bibfnamefont {J.}~\bibnamefont {Yu}}, \bibinfo {author}
  {\bibfnamefont {Y.}~\bibnamefont {Zhou}}, \bibinfo {author} {\bibfnamefont
  {V.}~\bibnamefont {Lordi}}, \ and\ \bibinfo {author} {\bibfnamefont
  {H.}~\bibnamefont {H\"affner}},\ }\href {\doibase 10.1103/PhysRevA.99.063427}
  {\bibfield  {journal} {\bibinfo  {journal} {Phys. Rev. A}\ }\textbf {\bibinfo
  {volume} {99}},\ \bibinfo {pages} {063427} (\bibinfo {year}
  {2019})}\BibitemShut {NoStop}%
\bibitem [{\citenamefont {Prajapati}\ \emph {et~al.}(2021)\citenamefont
  {Prajapati}, \citenamefont {Robinson}, \citenamefont {Berweger},
  \citenamefont {Simons}, \citenamefont {Artusio-Glimpse},\ and\ \citenamefont
  {Holloway}}]{Prajapati_2021}%
  \BibitemOpen
  \bibfield  {author} {\bibinfo {author} {\bibfnamefont {N.}~\bibnamefont
  {Prajapati}}, \bibinfo {author} {\bibfnamefont {A.~K.}\ \bibnamefont
  {Robinson}}, \bibinfo {author} {\bibfnamefont {S.}~\bibnamefont {Berweger}},
  \bibinfo {author} {\bibfnamefont {M.~T.}\ \bibnamefont {Simons}}, \bibinfo
  {author} {\bibfnamefont {A.~B.}\ \bibnamefont {Artusio-Glimpse}}, \ and\
  \bibinfo {author} {\bibfnamefont {C.~L.}\ \bibnamefont {Holloway}},\ }\href
  {\doibase 10.1063/5.0069195} {\bibfield  {journal} {\bibinfo  {journal}
  {Applied Physics Letters}\ }\textbf {\bibinfo {volume} {119}},\ \bibinfo
  {pages} {214001} (\bibinfo {year} {2021})}\BibitemShut {NoStop}%
\bibitem [{\citenamefont {Jing}\ \emph {et~al.}(2020)\citenamefont {Jing},
  \citenamefont {Hu}, \citenamefont {Ma}, \citenamefont {Zhang}, \citenamefont
  {Zhang}, \citenamefont {Xiao},\ and\ \citenamefont {Jia}}]{Jing_2020}%
  \BibitemOpen
  \bibfield  {author} {\bibinfo {author} {\bibfnamefont {M.}~\bibnamefont
  {Jing}}, \bibinfo {author} {\bibfnamefont {Y.}~\bibnamefont {Hu}}, \bibinfo
  {author} {\bibfnamefont {J.}~\bibnamefont {Ma}}, \bibinfo {author}
  {\bibfnamefont {H.}~\bibnamefont {Zhang}}, \bibinfo {author} {\bibfnamefont
  {L.}~\bibnamefont {Zhang}}, \bibinfo {author} {\bibfnamefont
  {L.}~\bibnamefont {Xiao}}, \ and\ \bibinfo {author} {\bibfnamefont
  {S.}~\bibnamefont {Jia}},\ }\href {\doibase 10.1038/s41567-020-0918-5}
  {\bibfield  {journal} {\bibinfo  {journal} {Nature Physics}\ }\textbf
  {\bibinfo {volume} {16}},\ \bibinfo {pages} {911} (\bibinfo {year}
  {2020})}\BibitemShut {NoStop}%
\bibitem [{\citenamefont {Ding}\ \emph {et~al.}(2020)\citenamefont {Ding},
  \citenamefont {Busche}, \citenamefont {Shi}, \citenamefont {Guo},\ and\
  \citenamefont {Adams}}]{Ding_2020}%
  \BibitemOpen
  \bibfield  {author} {\bibinfo {author} {\bibfnamefont {D.-S.}\ \bibnamefont
  {Ding}}, \bibinfo {author} {\bibfnamefont {H.}~\bibnamefont {Busche}},
  \bibinfo {author} {\bibfnamefont {B.-S.}\ \bibnamefont {Shi}}, \bibinfo
  {author} {\bibfnamefont {G.-C.}\ \bibnamefont {Guo}}, \ and\ \bibinfo
  {author} {\bibfnamefont {C.~S.}\ \bibnamefont {Adams}},\ }\href {\doibase
  10.1103/PhysRevX.10.021023} {\bibfield  {journal} {\bibinfo  {journal} {Phys.
  Rev. X}\ }\textbf {\bibinfo {volume} {10}},\ \bibinfo {pages} {021023}
  (\bibinfo {year} {2020})}\BibitemShut {NoStop}%
\bibitem [{\citenamefont {Carter}\ and\ \citenamefont
  {Martin}(2013)}]{Carter_2013}%
  \BibitemOpen
  \bibfield  {author} {\bibinfo {author} {\bibfnamefont {J.~D.}\ \bibnamefont
  {Carter}}\ and\ \bibinfo {author} {\bibfnamefont {J.~D.~D.}\ \bibnamefont
  {Martin}},\ }\href {\doibase 10.1103/PhysRevA.88.043429} {\bibfield
  {journal} {\bibinfo  {journal} {Phys. Rev. A}\ }\textbf {\bibinfo {volume}
  {88}},\ \bibinfo {pages} {043429} (\bibinfo {year} {2013})}\BibitemShut
  {NoStop}%
\bibitem [{\citenamefont {Shirotori}\ \emph {et~al.}(2021)\citenamefont
  {Shirotori}, \citenamefont {Kikitsu}, \citenamefont {Higashi}, \citenamefont
  {Kurosaki},\ and\ \citenamefont {Iwasaki}}]{Shirotori_2021}%
  \BibitemOpen
  \bibfield  {author} {\bibinfo {author} {\bibfnamefont {S.}~\bibnamefont
  {Shirotori}}, \bibinfo {author} {\bibfnamefont {A.}~\bibnamefont {Kikitsu}},
  \bibinfo {author} {\bibfnamefont {Y.}~\bibnamefont {Higashi}}, \bibinfo
  {author} {\bibfnamefont {Y.}~\bibnamefont {Kurosaki}}, \ and\ \bibinfo
  {author} {\bibfnamefont {H.}~\bibnamefont {Iwasaki}},\ }\href {\doibase
  10.1109/TMAG.2020.3012655} {\bibfield  {journal} {\bibinfo  {journal} {IEEE
  Transactions on Magnetics}\ }\textbf {\bibinfo {volume} {57}},\ \bibinfo
  {pages} {1} (\bibinfo {year} {2021})}\BibitemShut {NoStop}%
\bibitem [{\citenamefont {Hou}\ \emph {et~al.}(2022)\citenamefont {Hou},
  \citenamefont {Yuan}, \citenamefont {Wang}, \citenamefont {Itoh},\ and\
  \citenamefont {Maeda}}]{Hou_2022}%
  \BibitemOpen
  \bibfield  {author} {\bibinfo {author} {\bibfnamefont {Y.}~\bibnamefont
  {Hou}}, \bibinfo {author} {\bibfnamefont {W.}~\bibnamefont {Yuan}}, \bibinfo
  {author} {\bibfnamefont {D.~F.}\ \bibnamefont {Wang}}, \bibinfo {author}
  {\bibfnamefont {T.}~\bibnamefont {Itoh}}, \ and\ \bibinfo {author}
  {\bibfnamefont {R.}~\bibnamefont {Maeda}},\ }\href {\doibase
  10.1109/TIM.2021.3134315} {\bibfield  {journal} {\bibinfo  {journal} {IEEE
  Transactions on Instrumentation and Measurement}\ }\textbf {\bibinfo {volume}
  {71}},\ \bibinfo {pages} {1} (\bibinfo {year} {2022})}\BibitemShut {NoStop}%
\bibitem [{\citenamefont {Mercad\'e}\ \emph {et~al.}(2021)\citenamefont
  {Mercad\'e}, \citenamefont {Pelka}, \citenamefont {Burgwal}, \citenamefont
  {Xuereb}, \citenamefont {Mart\'{\i}nez},\ and\ \citenamefont
  {Verhagen}}]{Mercade_2021}%
  \BibitemOpen
  \bibfield  {author} {\bibinfo {author} {\bibfnamefont {L.}~\bibnamefont
  {Mercad\'e}}, \bibinfo {author} {\bibfnamefont {K.}~\bibnamefont {Pelka}},
  \bibinfo {author} {\bibfnamefont {R.}~\bibnamefont {Burgwal}}, \bibinfo
  {author} {\bibfnamefont {A.}~\bibnamefont {Xuereb}}, \bibinfo {author}
  {\bibfnamefont {A.}~\bibnamefont {Mart\'{\i}nez}}, \ and\ \bibinfo {author}
  {\bibfnamefont {E.}~\bibnamefont {Verhagen}},\ }\href {\doibase
  10.1103/PhysRevLett.127.073601} {\bibfield  {journal} {\bibinfo  {journal}
  {Phys. Rev. Lett.}\ }\textbf {\bibinfo {volume} {127}},\ \bibinfo {pages}
  {073601} (\bibinfo {year} {2021})}\BibitemShut {NoStop}%
\bibitem [{\citenamefont {Shandilya}\ \emph {et~al.}(2019)\citenamefont
  {Shandilya}, \citenamefont {Fröch}, \citenamefont {Mitchell}, \citenamefont
  {Lake}, \citenamefont {Kim}, \citenamefont {Toth}, \citenamefont {Behera},
  \citenamefont {Healey}, \citenamefont {Aharonovich},\ and\ \citenamefont
  {Barclay}}]{Shandilya_2019}%
  \BibitemOpen
  \bibfield  {author} {\bibinfo {author} {\bibfnamefont {P.~K.}\ \bibnamefont
  {Shandilya}}, \bibinfo {author} {\bibfnamefont {J.~E.}\ \bibnamefont
  {Fröch}}, \bibinfo {author} {\bibfnamefont {M.}~\bibnamefont {Mitchell}},
  \bibinfo {author} {\bibfnamefont {D.~P.}\ \bibnamefont {Lake}}, \bibinfo
  {author} {\bibfnamefont {S.}~\bibnamefont {Kim}}, \bibinfo {author}
  {\bibfnamefont {M.}~\bibnamefont {Toth}}, \bibinfo {author} {\bibfnamefont
  {B.}~\bibnamefont {Behera}}, \bibinfo {author} {\bibfnamefont
  {C.}~\bibnamefont {Healey}}, \bibinfo {author} {\bibfnamefont
  {I.}~\bibnamefont {Aharonovich}}, \ and\ \bibinfo {author} {\bibfnamefont
  {P.~E.}\ \bibnamefont {Barclay}},\ }\href {\doibase
  10.1021/acs.nanolett.8b04956} {\bibfield  {journal} {\bibinfo  {journal}
  {Nano Letters}\ }\textbf {\bibinfo {volume} {19}},\ \bibinfo {pages} {1343}
  (\bibinfo {year} {2019})},\ \bibinfo {note} {pMID: 30676758}\BibitemShut
  {NoStop}%
\bibitem [{\citenamefont {Zhang}\ \emph {et~al.}(2023)\citenamefont {Zhang},
  \citenamefont {Hines}, \citenamefont {Wilson},\ and\ \citenamefont
  {Guzman}}]{Zhang_2023}%
  \BibitemOpen
  \bibfield  {author} {\bibinfo {author} {\bibfnamefont {Y.}~\bibnamefont
  {Zhang}}, \bibinfo {author} {\bibfnamefont {A.}~\bibnamefont {Hines}},
  \bibinfo {author} {\bibfnamefont {D.~J.}\ \bibnamefont {Wilson}}, \ and\
  \bibinfo {author} {\bibfnamefont {F.}~\bibnamefont {Guzman}},\ }\href
  {\doibase 10.1103/PhysRevApplied.19.054004} {\bibfield  {journal} {\bibinfo
  {journal} {Phys. Rev. Appl.}\ }\textbf {\bibinfo {volume} {19}},\ \bibinfo
  {pages} {054004} (\bibinfo {year} {2023})}\BibitemShut {NoStop}%
\bibitem [{\citenamefont {DeVoe}\ and\ \citenamefont
  {Kurtsiefer}(2002)}]{DeVoe_2002}%
  \BibitemOpen
  \bibfield  {author} {\bibinfo {author} {\bibfnamefont {R.~G.}\ \bibnamefont
  {DeVoe}}\ and\ \bibinfo {author} {\bibfnamefont {C.}~\bibnamefont
  {Kurtsiefer}},\ }\href {\doibase 10.1103/PhysRevA.65.063407} {\bibfield
  {journal} {\bibinfo  {journal} {Phys. Rev. A}\ }\textbf {\bibinfo {volume}
  {65}},\ \bibinfo {pages} {063407} (\bibinfo {year} {2002})}\BibitemShut
  {NoStop}%
\bibitem [{\citenamefont {Porras}\ and\ \citenamefont
  {Cirac}(2004)}]{Porras_2004}%
  \BibitemOpen
  \bibfield  {author} {\bibinfo {author} {\bibfnamefont {D.}~\bibnamefont
  {Porras}}\ and\ \bibinfo {author} {\bibfnamefont {J.~I.}\ \bibnamefont
  {Cirac}},\ }\href {\doibase 10.1103/PhysRevLett.93.263602} {\bibfield
  {journal} {\bibinfo  {journal} {Phys. Rev. Lett.}\ }\textbf {\bibinfo
  {volume} {93}},\ \bibinfo {pages} {263602} (\bibinfo {year}
  {2004})}\BibitemShut {NoStop}%
\bibitem [{\citenamefont {Weber}\ \emph {et~al.}(2010)\citenamefont {Weber},
  \citenamefont {Fuhrer}, \citenamefont {Fasth}, \citenamefont {Lindwall},
  \citenamefont {Samuelson},\ and\ \citenamefont {Wacker}}]{Weber_2010}%
  \BibitemOpen
  \bibfield  {author} {\bibinfo {author} {\bibfnamefont {C.}~\bibnamefont
  {Weber}}, \bibinfo {author} {\bibfnamefont {A.}~\bibnamefont {Fuhrer}},
  \bibinfo {author} {\bibfnamefont {C.}~\bibnamefont {Fasth}}, \bibinfo
  {author} {\bibfnamefont {G.}~\bibnamefont {Lindwall}}, \bibinfo {author}
  {\bibfnamefont {L.}~\bibnamefont {Samuelson}}, \ and\ \bibinfo {author}
  {\bibfnamefont {A.}~\bibnamefont {Wacker}},\ }\href {\doibase
  10.1103/PhysRevLett.104.036801} {\bibfield  {journal} {\bibinfo  {journal}
  {Phys. Rev. Lett.}\ }\textbf {\bibinfo {volume} {104}},\ \bibinfo {pages}
  {036801} (\bibinfo {year} {2010})}\BibitemShut {NoStop}%
\bibitem [{\citenamefont {Brownnutt}\ \emph {et~al.}(2015)\citenamefont
  {Brownnutt}, \citenamefont {Kumph}, \citenamefont {Rabl},\ and\ \citenamefont
  {Blatt}}]{Brownnutt_2015}%
  \BibitemOpen
  \bibfield  {author} {\bibinfo {author} {\bibfnamefont {M.}~\bibnamefont
  {Brownnutt}}, \bibinfo {author} {\bibfnamefont {M.}~\bibnamefont {Kumph}},
  \bibinfo {author} {\bibfnamefont {P.}~\bibnamefont {Rabl}}, \ and\ \bibinfo
  {author} {\bibfnamefont {R.}~\bibnamefont {Blatt}},\ }\href {\doibase
  10.1103/RevModPhys.87.1419} {\bibfield  {journal} {\bibinfo  {journal} {Rev.
  Mod. Phys.}\ }\textbf {\bibinfo {volume} {87}},\ \bibinfo {pages} {1419}
  (\bibinfo {year} {2015})}\BibitemShut {NoStop}%
\bibitem [{\citenamefont {Hartke}\ \emph {et~al.}(2018)\citenamefont {Hartke},
  \citenamefont {Liu}, \citenamefont {Gullans},\ and\ \citenamefont
  {Petta}}]{Hartke_2018}%
  \BibitemOpen
  \bibfield  {author} {\bibinfo {author} {\bibfnamefont {T.~R.}\ \bibnamefont
  {Hartke}}, \bibinfo {author} {\bibfnamefont {Y.-Y.}\ \bibnamefont {Liu}},
  \bibinfo {author} {\bibfnamefont {M.~J.}\ \bibnamefont {Gullans}}, \ and\
  \bibinfo {author} {\bibfnamefont {J.~R.}\ \bibnamefont {Petta}},\ }\href
  {\doibase 10.1103/PhysRevLett.120.097701} {\bibfield  {journal} {\bibinfo
  {journal} {Phys. Rev. Lett.}\ }\textbf {\bibinfo {volume} {120}},\ \bibinfo
  {pages} {097701} (\bibinfo {year} {2018})}\BibitemShut {NoStop}%
\bibitem [{\citenamefont {Gullans}\ \emph {et~al.}(2018)\citenamefont
  {Gullans}, \citenamefont {Taylor},\ and\ \citenamefont
  {Petta}}]{Gullans_2018}%
  \BibitemOpen
  \bibfield  {author} {\bibinfo {author} {\bibfnamefont {M.~J.}\ \bibnamefont
  {Gullans}}, \bibinfo {author} {\bibfnamefont {J.~M.}\ \bibnamefont {Taylor}},
  \ and\ \bibinfo {author} {\bibfnamefont {J.~R.}\ \bibnamefont {Petta}},\
  }\href {\doibase 10.1103/PhysRevB.97.035305} {\bibfield  {journal} {\bibinfo
  {journal} {Phys. Rev. B}\ }\textbf {\bibinfo {volume} {97}},\ \bibinfo
  {pages} {035305} (\bibinfo {year} {2018})}\BibitemShut {NoStop}%
\bibitem [{\citenamefont {Catelani}\ and\ \citenamefont
  {Basko}(2019)}]{Catelani_2019}%
  \BibitemOpen
  \bibfield  {author} {\bibinfo {author} {\bibfnamefont {G.}~\bibnamefont
  {Catelani}}\ and\ \bibinfo {author} {\bibfnamefont {D.~M.}\ \bibnamefont
  {Basko}},\ }\href {\doibase 10.21468/SciPostPhys.6.1.013} {\bibfield
  {journal} {\bibinfo  {journal} {SciPost Phys.}\ }\textbf {\bibinfo {volume}
  {6}},\ \bibinfo {pages} {013} (\bibinfo {year} {2019})}\BibitemShut {NoStop}%
\bibitem [{\citenamefont {Wilen}\ \emph {et~al.}(2021)\citenamefont {Wilen},
  \citenamefont {Abdullah}, \citenamefont {Kurinsky}, \citenamefont {Stanford},
  \citenamefont {Cardani}, \citenamefont {D'Imperio}, \citenamefont {Tomei},
  \citenamefont {Faoro}, \citenamefont {Ioffe}, \citenamefont {Liu},
  \citenamefont {Opremcak}, \citenamefont {Christensen}, \citenamefont
  {DuBois},\ and\ \citenamefont {McDermott}}]{Wilen_2021}%
  \BibitemOpen
  \bibfield  {author} {\bibinfo {author} {\bibfnamefont {C.~D.}\ \bibnamefont
  {Wilen}}, \bibinfo {author} {\bibfnamefont {S.}~\bibnamefont {Abdullah}},
  \bibinfo {author} {\bibfnamefont {N.~A.}\ \bibnamefont {Kurinsky}}, \bibinfo
  {author} {\bibfnamefont {C.}~\bibnamefont {Stanford}}, \bibinfo {author}
  {\bibfnamefont {L.}~\bibnamefont {Cardani}}, \bibinfo {author} {\bibfnamefont
  {G.}~\bibnamefont {D'Imperio}}, \bibinfo {author} {\bibfnamefont
  {C.}~\bibnamefont {Tomei}}, \bibinfo {author} {\bibfnamefont
  {L.}~\bibnamefont {Faoro}}, \bibinfo {author} {\bibfnamefont {L.~B.}\
  \bibnamefont {Ioffe}}, \bibinfo {author} {\bibfnamefont {C.~H.}\ \bibnamefont
  {Liu}}, \bibinfo {author} {\bibfnamefont {A.}~\bibnamefont {Opremcak}},
  \bibinfo {author} {\bibfnamefont {B.~G.}\ \bibnamefont {Christensen}},
  \bibinfo {author} {\bibfnamefont {J.~L.}\ \bibnamefont {DuBois}}, \ and\
  \bibinfo {author} {\bibfnamefont {R.}~\bibnamefont {McDermott}},\ }\href
  {\doibase 10.1038/s41586-021-03557-5} {\bibfield  {journal} {\bibinfo
  {journal} {Nature}\ }\textbf {\bibinfo {volume} {594}},\ \bibinfo {pages}
  {369} (\bibinfo {year} {2021})}\BibitemShut {NoStop}%
\bibitem [{\citenamefont {Siddiqi}(2021)}]{Siddiqi_2021}%
  \BibitemOpen
  \bibfield  {author} {\bibinfo {author} {\bibfnamefont {I.}~\bibnamefont
  {Siddiqi}},\ }\href {\doibase 10.1038/s41578-021-00370-4} {\bibfield
  {journal} {\bibinfo  {journal} {Nature Reviews Materials}\ }\textbf {\bibinfo
  {volume} {6}},\ \bibinfo {pages} {875} (\bibinfo {year} {2021})}\BibitemShut
  {NoStop}%
\bibitem [{\citenamefont {Iaia}\ \emph {et~al.}(2022)\citenamefont {Iaia},
  \citenamefont {Ku}, \citenamefont {Ballard}, \citenamefont {Larson},
  \citenamefont {Yelton}, \citenamefont {Liu}, \citenamefont {Patel},
  \citenamefont {McDermott},\ and\ \citenamefont {Plourde}}]{Iaia_2022}%
  \BibitemOpen
  \bibfield  {author} {\bibinfo {author} {\bibfnamefont {V.}~\bibnamefont
  {Iaia}}, \bibinfo {author} {\bibfnamefont {J.}~\bibnamefont {Ku}}, \bibinfo
  {author} {\bibfnamefont {A.}~\bibnamefont {Ballard}}, \bibinfo {author}
  {\bibfnamefont {C.~P.}\ \bibnamefont {Larson}}, \bibinfo {author}
  {\bibfnamefont {E.}~\bibnamefont {Yelton}}, \bibinfo {author} {\bibfnamefont
  {C.~H.}\ \bibnamefont {Liu}}, \bibinfo {author} {\bibfnamefont
  {S.}~\bibnamefont {Patel}}, \bibinfo {author} {\bibfnamefont
  {R.}~\bibnamefont {McDermott}}, \ and\ \bibinfo {author} {\bibfnamefont
  {B.~L.~T.}\ \bibnamefont {Plourde}},\ }\href {\doibase
  10.1038/s41467-022-33997-0} {\bibfield  {journal} {\bibinfo  {journal}
  {Nature Communications}\ }\textbf {\bibinfo {volume} {13}},\ \bibinfo {pages}
  {6425} (\bibinfo {year} {2022})}\BibitemShut {NoStop}%
\bibitem [{\citenamefont {Bargerbos}\ \emph {et~al.}(2023)\citenamefont
  {Bargerbos}, \citenamefont {Splitthoff}, \citenamefont {Pita-Vidal},
  \citenamefont {Wesdorp}, \citenamefont {Liu}, \citenamefont {Krogstrup},
  \citenamefont {Kouwenhoven}, \citenamefont {Andersen},\ and\ \citenamefont
  {Gr\"unhaupt}}]{Bargerbos_2023}%
  \BibitemOpen
  \bibfield  {author} {\bibinfo {author} {\bibfnamefont {A.}~\bibnamefont
  {Bargerbos}}, \bibinfo {author} {\bibfnamefont {L.~J.}\ \bibnamefont
  {Splitthoff}}, \bibinfo {author} {\bibfnamefont {M.}~\bibnamefont
  {Pita-Vidal}}, \bibinfo {author} {\bibfnamefont {J.~J.}\ \bibnamefont
  {Wesdorp}}, \bibinfo {author} {\bibfnamefont {Y.}~\bibnamefont {Liu}},
  \bibinfo {author} {\bibfnamefont {P.}~\bibnamefont {Krogstrup}}, \bibinfo
  {author} {\bibfnamefont {L.~P.}\ \bibnamefont {Kouwenhoven}}, \bibinfo
  {author} {\bibfnamefont {C.~K.}\ \bibnamefont {Andersen}}, \ and\ \bibinfo
  {author} {\bibfnamefont {L.}~\bibnamefont {Gr\"unhaupt}},\ }\href {\doibase
  10.1103/PhysRevApplied.19.024014} {\bibfield  {journal} {\bibinfo  {journal}
  {Phys. Rev. Appl.}\ }\textbf {\bibinfo {volume} {19}},\ \bibinfo {pages}
  {024014} (\bibinfo {year} {2023})}\BibitemShut {NoStop}%
\bibitem [{\citenamefont {Freitag}\ \emph {et~al.}(2017)\citenamefont
  {Freitag}, \citenamefont {Heck\"otter}, \citenamefont {Bayer},\ and\
  \citenamefont {A\ss{}mann}}]{Freitag_2017}%
  \BibitemOpen
  \bibfield  {author} {\bibinfo {author} {\bibfnamefont {M.}~\bibnamefont
  {Freitag}}, \bibinfo {author} {\bibfnamefont {J.}~\bibnamefont
  {Heck\"otter}}, \bibinfo {author} {\bibfnamefont {M.}~\bibnamefont {Bayer}},
  \ and\ \bibinfo {author} {\bibfnamefont {M.}~\bibnamefont {A\ss{}mann}},\
  }\href {\doibase 10.1103/PhysRevB.95.155204} {\bibfield  {journal} {\bibinfo
  {journal} {Phys. Rev. B}\ }\textbf {\bibinfo {volume} {95}},\ \bibinfo
  {pages} {155204} (\bibinfo {year} {2017})}\BibitemShut {NoStop}%
\bibitem [{\citenamefont {Mazza}\ \emph {et~al.}(2020)\citenamefont {Mazza},
  \citenamefont {Schmidt},\ and\ \citenamefont {Lesanovsky}}]{Mazza_2020}%
  \BibitemOpen
  \bibfield  {author} {\bibinfo {author} {\bibfnamefont {P.~P.}\ \bibnamefont
  {Mazza}}, \bibinfo {author} {\bibfnamefont {R.}~\bibnamefont {Schmidt}}, \
  and\ \bibinfo {author} {\bibfnamefont {I.}~\bibnamefont {Lesanovsky}},\
  }\href {\doibase 10.1103/PhysRevLett.125.033602} {\bibfield  {journal}
  {\bibinfo  {journal} {Phys. Rev. Lett.}\ }\textbf {\bibinfo {volume} {125}},\
  \bibinfo {pages} {033602} (\bibinfo {year} {2020})}\BibitemShut {NoStop}%
\bibitem [{\citenamefont {Teoh}\ \emph {et~al.}(2021)\citenamefont {Teoh},
  \citenamefont {Sajjan}, \citenamefont {Sun}, \citenamefont {Rajabi},\ and\
  \citenamefont {Islam}}]{Teoh_2021}%
  \BibitemOpen
  \bibfield  {author} {\bibinfo {author} {\bibfnamefont {Y.~H.}\ \bibnamefont
  {Teoh}}, \bibinfo {author} {\bibfnamefont {M.}~\bibnamefont {Sajjan}},
  \bibinfo {author} {\bibfnamefont {Z.}~\bibnamefont {Sun}}, \bibinfo {author}
  {\bibfnamefont {F.}~\bibnamefont {Rajabi}}, \ and\ \bibinfo {author}
  {\bibfnamefont {R.}~\bibnamefont {Islam}},\ }\href {\doibase
  10.1103/PhysRevA.104.022420} {\bibfield  {journal} {\bibinfo  {journal}
  {Phys. Rev. A}\ }\textbf {\bibinfo {volume} {104}},\ \bibinfo {pages}
  {022420} (\bibinfo {year} {2021})}\BibitemShut {NoStop}%
\bibitem [{\citenamefont {Romaszko}\ \emph {et~al.}(2020)\citenamefont
  {Romaszko}, \citenamefont {Hong}, \citenamefont {Siegele}, \citenamefont
  {Puddy}, \citenamefont {Lebrun-Gallagher}, \citenamefont {Weidt},\ and\
  \citenamefont {Hensinger}}]{Romaszko_2020}%
  \BibitemOpen
  \bibfield  {author} {\bibinfo {author} {\bibfnamefont {Z.~D.}\ \bibnamefont
  {Romaszko}}, \bibinfo {author} {\bibfnamefont {S.}~\bibnamefont {Hong}},
  \bibinfo {author} {\bibfnamefont {M.}~\bibnamefont {Siegele}}, \bibinfo
  {author} {\bibfnamefont {R.~K.}\ \bibnamefont {Puddy}}, \bibinfo {author}
  {\bibfnamefont {F.~R.}\ \bibnamefont {Lebrun-Gallagher}}, \bibinfo {author}
  {\bibfnamefont {S.}~\bibnamefont {Weidt}}, \ and\ \bibinfo {author}
  {\bibfnamefont {W.~K.}\ \bibnamefont {Hensinger}},\ }\href {\doibase
  10.1038/s42254-020-0182-8} {\bibfield  {journal} {\bibinfo  {journal} {Nature
  Reviews Physics}\ }\textbf {\bibinfo {volume} {2}},\ \bibinfo {pages} {285}
  (\bibinfo {year} {2020})}\BibitemShut {NoStop}%
\bibitem [{\citenamefont {Goel}\ and\ \citenamefont
  {Freericks}(2021)}]{Goel_2021}%
  \BibitemOpen
  \bibfield  {author} {\bibinfo {author} {\bibfnamefont {N.}~\bibnamefont
  {Goel}}\ and\ \bibinfo {author} {\bibfnamefont {J.~K.}\ \bibnamefont
  {Freericks}},\ }\href {\doibase 10.1088/2058-9565/ac1e02} {\bibfield
  {journal} {\bibinfo  {journal} {Quantum Science and Technology}\ }\textbf
  {\bibinfo {volume} {6}},\ \bibinfo {pages} {044010} (\bibinfo {year}
  {2021})}\BibitemShut {NoStop}%
\bibitem [{\citenamefont {Clark}\ \emph {et~al.}(2021)\citenamefont {Clark},
  \citenamefont {Chekhova}, \citenamefont {Matthews}, \citenamefont {Rarity},\
  and\ \citenamefont {Oulton}}]{Clark_2021}%
  \BibitemOpen
  \bibfield  {author} {\bibinfo {author} {\bibfnamefont {A.~S.}\ \bibnamefont
  {Clark}}, \bibinfo {author} {\bibfnamefont {M.}~\bibnamefont {Chekhova}},
  \bibinfo {author} {\bibfnamefont {J.~C.~F.}\ \bibnamefont {Matthews}},
  \bibinfo {author} {\bibfnamefont {J.~G.}\ \bibnamefont {Rarity}}, \ and\
  \bibinfo {author} {\bibfnamefont {R.~F.}\ \bibnamefont {Oulton}},\ }\href
  {\doibase 10.1063/5.0041043} {\bibfield  {journal} {\bibinfo  {journal}
  {Applied Physics Letters}\ }\textbf {\bibinfo {volume} {118}} (\bibinfo
  {year} {2021}),\ 10.1063/5.0041043},\ \bibinfo {note} {060401}\BibitemShut
  {NoStop}%
\bibitem [{\citenamefont {Ruo-Berchera}\ \emph {et~al.}(2020)\citenamefont
  {Ruo-Berchera}, \citenamefont {Meda}, \citenamefont {Losero}, \citenamefont
  {Avella}, \citenamefont {Samantaray},\ and\ \citenamefont
  {Genovese}}]{Ruo-Berchera_2020}%
  \BibitemOpen
  \bibfield  {author} {\bibinfo {author} {\bibfnamefont {I.}~\bibnamefont
  {Ruo-Berchera}}, \bibinfo {author} {\bibfnamefont {A.}~\bibnamefont {Meda}},
  \bibinfo {author} {\bibfnamefont {E.}~\bibnamefont {Losero}}, \bibinfo
  {author} {\bibfnamefont {A.}~\bibnamefont {Avella}}, \bibinfo {author}
  {\bibfnamefont {N.}~\bibnamefont {Samantaray}}, \ and\ \bibinfo {author}
  {\bibfnamefont {M.}~\bibnamefont {Genovese}},\ }\href {\doibase
  10.1063/5.0009538} {\bibfield  {journal} {\bibinfo  {journal} {Applied
  Physics Letters}\ }\textbf {\bibinfo {volume} {116}} (\bibinfo {year}
  {2020}),\ 10.1063/5.0009538},\ \bibinfo {note} {214001}\BibitemShut {NoStop}%
\bibitem [{\citenamefont {Tan}\ and\ \citenamefont {Jeong}(2019)}]{Tan_2019}%
  \BibitemOpen
  \bibfield  {author} {\bibinfo {author} {\bibfnamefont {K.~C.}\ \bibnamefont
  {Tan}}\ and\ \bibinfo {author} {\bibfnamefont {H.}~\bibnamefont {Jeong}},\
  }\href {\doibase 10.1116/1.5126696} {\bibfield  {journal} {\bibinfo
  {journal} {AVS Quantum Science}\ }\textbf {\bibinfo {volume} {1}} (\bibinfo
  {year} {2019}),\ 10.1116/1.5126696},\ \bibinfo {note} {014701}\BibitemShut
  {NoStop}%
\bibitem [{\citenamefont {Yadin}\ \emph {et~al.}(2018)\citenamefont {Yadin},
  \citenamefont {Binder}, \citenamefont {Thompson}, \citenamefont
  {Narasimhachar}, \citenamefont {Gu},\ and\ \citenamefont {Kim}}]{Yadin_2018}%
  \BibitemOpen
  \bibfield  {author} {\bibinfo {author} {\bibfnamefont {B.}~\bibnamefont
  {Yadin}}, \bibinfo {author} {\bibfnamefont {F.~C.}\ \bibnamefont {Binder}},
  \bibinfo {author} {\bibfnamefont {J.}~\bibnamefont {Thompson}}, \bibinfo
  {author} {\bibfnamefont {V.}~\bibnamefont {Narasimhachar}}, \bibinfo {author}
  {\bibfnamefont {M.}~\bibnamefont {Gu}}, \ and\ \bibinfo {author}
  {\bibfnamefont {M.~S.}\ \bibnamefont {Kim}},\ }\href {\doibase
  10.1103/PhysRevX.8.041038} {\bibfield  {journal} {\bibinfo  {journal} {Phys.
  Rev. X}\ }\textbf {\bibinfo {volume} {8}},\ \bibinfo {pages} {041038}
  (\bibinfo {year} {2018})}\BibitemShut {NoStop}%
\bibitem [{\citenamefont {Moreau}\ \emph {et~al.}(2017)\citenamefont {Moreau},
  \citenamefont {Sabines-Chesterking}, \citenamefont {Whittaker}, \citenamefont
  {Joshi}, \citenamefont {Birchall}, \citenamefont {McMillan}, \citenamefont
  {Rarity},\ and\ \citenamefont {Matthews}}]{Moreau_2017}%
  \BibitemOpen
  \bibfield  {author} {\bibinfo {author} {\bibfnamefont {P.-A.}\ \bibnamefont
  {Moreau}}, \bibinfo {author} {\bibfnamefont {J.}~\bibnamefont
  {Sabines-Chesterking}}, \bibinfo {author} {\bibfnamefont {R.}~\bibnamefont
  {Whittaker}}, \bibinfo {author} {\bibfnamefont {S.~K.}\ \bibnamefont
  {Joshi}}, \bibinfo {author} {\bibfnamefont {P.~M.}\ \bibnamefont {Birchall}},
  \bibinfo {author} {\bibfnamefont {A.}~\bibnamefont {McMillan}}, \bibinfo
  {author} {\bibfnamefont {J.~G.}\ \bibnamefont {Rarity}}, \ and\ \bibinfo
  {author} {\bibfnamefont {J.~C.~F.}\ \bibnamefont {Matthews}},\ }\href
  {\doibase 10.1038/s41598-017-06545-w} {\bibfield  {journal} {\bibinfo
  {journal} {Scientific Reports}\ }\textbf {\bibinfo {volume} {7}},\ \bibinfo
  {pages} {6256} (\bibinfo {year} {2017})}\BibitemShut {NoStop}%
\bibitem [{\citenamefont {Brida}\ \emph {et~al.}(2010)\citenamefont {Brida},
  \citenamefont {Genovese},\ and\ \citenamefont {Ruo~Berchera}}]{Brida_2010}%
  \BibitemOpen
  \bibfield  {author} {\bibinfo {author} {\bibfnamefont {G.}~\bibnamefont
  {Brida}}, \bibinfo {author} {\bibfnamefont {M.}~\bibnamefont {Genovese}}, \
  and\ \bibinfo {author} {\bibfnamefont {I.}~\bibnamefont {Ruo~Berchera}},\
  }\href {\doibase 10.1038/nphoton.2010.29} {\bibfield  {journal} {\bibinfo
  {journal} {Nature Photonics}\ }\textbf {\bibinfo {volume} {4}},\ \bibinfo
  {pages} {227} (\bibinfo {year} {2010})}\BibitemShut {NoStop}%
\bibitem [{\citenamefont {Rivas}\ and\ \citenamefont
  {Luis}(2010)}]{Rivas_2010}%
  \BibitemOpen
  \bibfield  {author} {\bibinfo {author} {\bibfnamefont {A.}~\bibnamefont
  {Rivas}}\ and\ \bibinfo {author} {\bibfnamefont {A.}~\bibnamefont {Luis}},\
  }\href {\doibase 10.1103/PhysRevLett.105.010403} {\bibfield  {journal}
  {\bibinfo  {journal} {Phys. Rev. Lett.}\ }\textbf {\bibinfo {volume} {105}},\
  \bibinfo {pages} {010403} (\bibinfo {year} {2010})}\BibitemShut {NoStop}%
\bibitem [{\citenamefont {Bhattacharjee}\ and\ \citenamefont
  {Dutta}(2021)}]{Bhattacharjee_2021}%
  \BibitemOpen
  \bibfield  {author} {\bibinfo {author} {\bibfnamefont {S.}~\bibnamefont
  {Bhattacharjee}}\ and\ \bibinfo {author} {\bibfnamefont {A.}~\bibnamefont
  {Dutta}},\ }\href {\doibase 10.1140/epjb/s10051-021-00235-3} {\bibfield
  {journal} {\bibinfo  {journal} {The European Physical Journal B}\ }\textbf
  {\bibinfo {volume} {94}},\ \bibinfo {pages} {239} (\bibinfo {year}
  {2021})}\BibitemShut {NoStop}%
\bibitem [{\citenamefont {Myers}\ \emph {et~al.}(2022)\citenamefont {Myers},
  \citenamefont {Abah},\ and\ \citenamefont {Deffner}}]{Myers_2022}%
  \BibitemOpen
  \bibfield  {author} {\bibinfo {author} {\bibfnamefont {N.~M.}\ \bibnamefont
  {Myers}}, \bibinfo {author} {\bibfnamefont {O.}~\bibnamefont {Abah}}, \ and\
  \bibinfo {author} {\bibfnamefont {S.}~\bibnamefont {Deffner}},\ }\href
  {\doibase 10.1116/5.0083192} {\bibfield  {journal} {\bibinfo  {journal} {AVS
  Quantum Science}\ }\textbf {\bibinfo {volume} {4}},\ \bibinfo {pages}
  {027101} (\bibinfo {year} {2022})}\BibitemShut {NoStop}%
\bibitem [{\citenamefont {Cangemi}\ \emph {et~al.}(2023)\citenamefont
  {Cangemi}, \citenamefont {Bhadra},\ and\ \citenamefont
  {Levy}}]{Cangemi_2023}%
  \BibitemOpen
  \bibfield  {author} {\bibinfo {author} {\bibfnamefont {L.~M.}\ \bibnamefont
  {Cangemi}}, \bibinfo {author} {\bibfnamefont {C.}~\bibnamefont {Bhadra}}, \
  and\ \bibinfo {author} {\bibfnamefont {A.}~\bibnamefont {Levy}},\ }\href@noop
  {} {\  (\bibinfo {year} {2023})},\ \Eprint {http://arxiv.org/abs/2302.00726}
  {arXiv:2302.00726 [quant-ph]} \BibitemShut {NoStop}%
\bibitem [{\citenamefont {Bloch}(1957)}]{Bloch_1957}%
  \BibitemOpen
  \bibfield  {author} {\bibinfo {author} {\bibfnamefont {F.}~\bibnamefont
  {Bloch}},\ }\href {\doibase 10.1103/PhysRev.105.1206} {\bibfield  {journal}
  {\bibinfo  {journal} {Phys. Rev.}\ }\textbf {\bibinfo {volume} {105}},\
  \bibinfo {pages} {1206} (\bibinfo {year} {1957})}\BibitemShut {NoStop}%
\bibitem [{\citenamefont {Redfield}(1957)}]{Redfield_1957}%
  \BibitemOpen
  \bibfield  {author} {\bibinfo {author} {\bibfnamefont {A.~G.}\ \bibnamefont
  {Redfield}},\ }\href {\doibase 10.1147/rd.11.0019} {\bibfield  {journal}
  {\bibinfo  {journal} {IBM Journal of Research and Development}\ }\textbf
  {\bibinfo {volume} {1}},\ \bibinfo {pages} {19} (\bibinfo {year}
  {1957})}\BibitemShut {NoStop}%
\bibitem [{\citenamefont {Breuer}\ and\ \citenamefont
  {Petruccione}(2006)}]{Breuer_book}%
  \BibitemOpen
  \bibfield  {author} {\bibinfo {author} {\bibfnamefont {H.-P.}\ \bibnamefont
  {Breuer}}\ and\ \bibinfo {author} {\bibfnamefont {F.}~\bibnamefont
  {Petruccione}},\ }\href@noop {} {\emph {\bibinfo {title} {The Theory of Open
  Quantum Systems}}}\ (\bibinfo  {publisher} {Oxford University Press,
  Oxford},\ \bibinfo {year} {2006})\BibitemShut {NoStop}%
\bibitem [{\citenamefont {Tupkary}\ \emph {et~al.}(2023)\citenamefont
  {Tupkary}, \citenamefont {Dhar}, \citenamefont {Kulkarni},\ and\
  \citenamefont {Purkayastha}}]{Tupkary_2023}%
  \BibitemOpen
  \bibfield  {author} {\bibinfo {author} {\bibfnamefont {D.}~\bibnamefont
  {Tupkary}}, \bibinfo {author} {\bibfnamefont {A.}~\bibnamefont {Dhar}},
  \bibinfo {author} {\bibfnamefont {M.}~\bibnamefont {Kulkarni}}, \ and\
  \bibinfo {author} {\bibfnamefont {A.}~\bibnamefont {Purkayastha}},\ }\href
  {\doibase 10.1103/PhysRevA.107.062216} {\bibfield  {journal} {\bibinfo
  {journal} {Phys. Rev. A}\ }\textbf {\bibinfo {volume} {107}},\ \bibinfo
  {pages} {062216} (\bibinfo {year} {2023})}\BibitemShut {NoStop}%
\bibitem [{\citenamefont {Tupkary}\ \emph {et~al.}(2022)\citenamefont
  {Tupkary}, \citenamefont {Dhar}, \citenamefont {Kulkarni},\ and\
  \citenamefont {Purkayastha}}]{Tupkary_2022}%
  \BibitemOpen
  \bibfield  {author} {\bibinfo {author} {\bibfnamefont {D.}~\bibnamefont
  {Tupkary}}, \bibinfo {author} {\bibfnamefont {A.}~\bibnamefont {Dhar}},
  \bibinfo {author} {\bibfnamefont {M.}~\bibnamefont {Kulkarni}}, \ and\
  \bibinfo {author} {\bibfnamefont {A.}~\bibnamefont {Purkayastha}},\ }\href
  {\doibase 10.1103/PhysRevA.105.032208} {\bibfield  {journal} {\bibinfo
  {journal} {Phys. Rev. A}\ }\textbf {\bibinfo {volume} {105}},\ \bibinfo
  {pages} {032208} (\bibinfo {year} {2022})}\BibitemShut {NoStop}%
\bibitem [{\citenamefont {Hartmann}\ and\ \citenamefont
  {Strunz}(2020)}]{Hartmann_2020}%
  \BibitemOpen
  \bibfield  {author} {\bibinfo {author} {\bibfnamefont {R.}~\bibnamefont
  {Hartmann}}\ and\ \bibinfo {author} {\bibfnamefont {W.~T.}\ \bibnamefont
  {Strunz}},\ }\href {\doibase 10.1103/PhysRevA.101.012103} {\bibfield
  {journal} {\bibinfo  {journal} {Phys. Rev. A}\ }\textbf {\bibinfo {volume}
  {101}},\ \bibinfo {pages} {012103} (\bibinfo {year} {2020})}\BibitemShut
  {NoStop}%
\bibitem [{\citenamefont {Fleming}\ and\ \citenamefont
  {Cummings}(2011)}]{Fleming_2011}%
  \BibitemOpen
  \bibfield  {author} {\bibinfo {author} {\bibfnamefont {C.~H.}\ \bibnamefont
  {Fleming}}\ and\ \bibinfo {author} {\bibfnamefont {N.~I.}\ \bibnamefont
  {Cummings}},\ }\href {\doibase 10.1103/PhysRevE.83.031117} {\bibfield
  {journal} {\bibinfo  {journal} {Phys. Rev. E}\ }\textbf {\bibinfo {volume}
  {83}},\ \bibinfo {pages} {031117} (\bibinfo {year} {2011})}\BibitemShut
  {NoStop}%
\bibitem [{\citenamefont {Khan}\ \emph {et~al.}(2023)\citenamefont {Khan},
  \citenamefont {Agarwalla},\ and\ \citenamefont {Jain}}]{Khan_2023}%
  \BibitemOpen
  \bibfield  {author} {\bibinfo {author} {\bibfnamefont {S.}~\bibnamefont
  {Khan}}, \bibinfo {author} {\bibfnamefont {B.~K.}\ \bibnamefont {Agarwalla}},
  \ and\ \bibinfo {author} {\bibfnamefont {S.}~\bibnamefont {Jain}},\
  }\href@noop {} {\  (\bibinfo {year} {2023})},\ \Eprint
  {http://arxiv.org/abs/2306.04677} {arXiv:2306.04677 [quant-ph]} \BibitemShut
  {NoStop}%
\bibitem [{\citenamefont {Kjaergaard}\ \emph {et~al.}(2020)\citenamefont
  {Kjaergaard}, \citenamefont {Schwartz}, \citenamefont {Braum\"{u}ller},
  \citenamefont {Krantz}, \citenamefont {Wang}, \citenamefont {Gustavsson},\
  and\ \citenamefont {Oliver}}]{Kjaergaard_2020}%
  \BibitemOpen
  \bibfield  {author} {\bibinfo {author} {\bibfnamefont {M.}~\bibnamefont
  {Kjaergaard}}, \bibinfo {author} {\bibfnamefont {M.~E.}\ \bibnamefont
  {Schwartz}}, \bibinfo {author} {\bibfnamefont {J.}~\bibnamefont
  {Braum\"{u}ller}}, \bibinfo {author} {\bibfnamefont {P.}~\bibnamefont
  {Krantz}}, \bibinfo {author} {\bibfnamefont {J.~I.-J.}\ \bibnamefont {Wang}},
  \bibinfo {author} {\bibfnamefont {S.}~\bibnamefont {Gustavsson}}, \ and\
  \bibinfo {author} {\bibfnamefont {W.~D.}\ \bibnamefont {Oliver}},\ }\href
  {\doibase 10.1146/annurev-conmatphys-031119-050605} {\bibfield  {journal}
  {\bibinfo  {journal} {Annual Review of Condensed Matter Physics}\ }\textbf
  {\bibinfo {volume} {11}},\ \bibinfo {pages} {369} (\bibinfo {year}
  {2020})}\BibitemShut {NoStop}%
\bibitem [{\citenamefont {Dmitriev}\ and\ \citenamefont
  {Astafiev}(2021)}]{Dmitriev_2021}%
  \BibitemOpen
  \bibfield  {author} {\bibinfo {author} {\bibfnamefont {A.~Y.}\ \bibnamefont
  {Dmitriev}}\ and\ \bibinfo {author} {\bibfnamefont {O.}~\bibnamefont
  {Astafiev}},\ }\href@noop {} {\bibfield  {journal} {\bibinfo  {journal}
  {Applied Physics Letters}\ }\textbf {\bibinfo {volume} {119}},\ \bibinfo
  {pages} {080501} (\bibinfo {year} {2021})}\BibitemShut {NoStop}%
\bibitem [{\citenamefont {Purkayastha}\ \emph {et~al.}(2016)\citenamefont
  {Purkayastha}, \citenamefont {Dhar},\ and\ \citenamefont
  {Kulkarni}}]{Purkayastha_2016}%
  \BibitemOpen
  \bibfield  {author} {\bibinfo {author} {\bibfnamefont {A.}~\bibnamefont
  {Purkayastha}}, \bibinfo {author} {\bibfnamefont {A.}~\bibnamefont {Dhar}}, \
  and\ \bibinfo {author} {\bibfnamefont {M.}~\bibnamefont {Kulkarni}},\ }\href
  {\doibase 10.1103/PhysRevA.94.052134} {\bibfield  {journal} {\bibinfo
  {journal} {Phys. Rev. A}\ }\textbf {\bibinfo {volume} {94}},\ \bibinfo
  {pages} {052134} (\bibinfo {year} {2016})}\BibitemShut {NoStop}%
\bibitem [{\citenamefont {Walls}\ and\ \citenamefont
  {Milburn}(2008)}]{Walls_book}%
  \BibitemOpen
  \bibfield  {author} {\bibinfo {author} {\bibfnamefont {D.~F.}\ \bibnamefont
  {Walls}}\ and\ \bibinfo {author} {\bibfnamefont {G.~J.}\ \bibnamefont
  {Milburn}},\ }\href@noop {} {\emph {\bibinfo {title} {Quantum Optics}}}\
  (\bibinfo  {publisher} {Springer-Verlag Berlin, Heidelberg},\ \bibinfo {year}
  {2008})\BibitemShut {NoStop}%
\bibitem [{\citenamefont {Kewming}\ and\ \citenamefont
  {Shrapnel}(2022)}]{Kewming_2022}%
  \BibitemOpen
  \bibfield  {author} {\bibinfo {author} {\bibfnamefont {M.~J.}\ \bibnamefont
  {Kewming}}\ and\ \bibinfo {author} {\bibfnamefont {S.}~\bibnamefont
  {Shrapnel}},\ }\href {\doibase 10.22331/q-2022-04-13-685} {\bibfield
  {journal} {\bibinfo  {journal} {Quantum}\ }\textbf {\bibinfo {volume} {6}},\
  \bibinfo {pages} {685} (\bibinfo {year} {2022})}\BibitemShut {NoStop}%
\bibitem [{\citenamefont {Ge}\ \emph {et~al.}(2019)\citenamefont {Ge},
  \citenamefont {Sawyer}, \citenamefont {Britton}, \citenamefont {Jacobs},
  \citenamefont {Bollinger},\ and\ \citenamefont {Foss-Feig}}]{Ge_2019}%
  \BibitemOpen
  \bibfield  {author} {\bibinfo {author} {\bibfnamefont {W.}~\bibnamefont
  {Ge}}, \bibinfo {author} {\bibfnamefont {B.~C.}\ \bibnamefont {Sawyer}},
  \bibinfo {author} {\bibfnamefont {J.~W.}\ \bibnamefont {Britton}}, \bibinfo
  {author} {\bibfnamefont {K.}~\bibnamefont {Jacobs}}, \bibinfo {author}
  {\bibfnamefont {J.~J.}\ \bibnamefont {Bollinger}}, \ and\ \bibinfo {author}
  {\bibfnamefont {M.}~\bibnamefont {Foss-Feig}},\ }\href {\doibase
  10.1103/PhysRevLett.122.030501} {\bibfield  {journal} {\bibinfo  {journal}
  {Phys. Rev. Lett.}\ }\textbf {\bibinfo {volume} {122}},\ \bibinfo {pages}
  {030501} (\bibinfo {year} {2019})}\BibitemShut {NoStop}%
\bibitem [{\citenamefont {Gambetta}\ \emph {et~al.}(2020)\citenamefont
  {Gambetta}, \citenamefont {Li}, \citenamefont {Schmidt-Kaler},\ and\
  \citenamefont {Lesanovsky}}]{Gambetta_2020}%
  \BibitemOpen
  \bibfield  {author} {\bibinfo {author} {\bibfnamefont {F.~M.}\ \bibnamefont
  {Gambetta}}, \bibinfo {author} {\bibfnamefont {W.}~\bibnamefont {Li}},
  \bibinfo {author} {\bibfnamefont {F.}~\bibnamefont {Schmidt-Kaler}}, \ and\
  \bibinfo {author} {\bibfnamefont {I.}~\bibnamefont {Lesanovsky}},\ }\href
  {\doibase 10.1103/PhysRevLett.124.043402} {\bibfield  {journal} {\bibinfo
  {journal} {Phys. Rev. Lett.}\ }\textbf {\bibinfo {volume} {124}},\ \bibinfo
  {pages} {043402} (\bibinfo {year} {2020})}\BibitemShut {NoStop}%
\bibitem [{\citenamefont {Davoudi}\ \emph {et~al.}(2021)\citenamefont
  {Davoudi}, \citenamefont {Linke},\ and\ \citenamefont
  {Pagano}}]{Davoudi_2021}%
  \BibitemOpen
  \bibfield  {author} {\bibinfo {author} {\bibfnamefont {Z.}~\bibnamefont
  {Davoudi}}, \bibinfo {author} {\bibfnamefont {N.~M.}\ \bibnamefont {Linke}},
  \ and\ \bibinfo {author} {\bibfnamefont {G.}~\bibnamefont {Pagano}},\ }\href
  {\doibase 10.1103/PhysRevResearch.3.043072} {\bibfield  {journal} {\bibinfo
  {journal} {Phys. Rev. Res.}\ }\textbf {\bibinfo {volume} {3}},\ \bibinfo
  {pages} {043072} (\bibinfo {year} {2021})}\BibitemShut {NoStop}%
\bibitem [{\citenamefont {Tao}\ \emph {et~al.}(2022)\citenamefont {Tao},
  \citenamefont {Zhang}, \citenamefont {Zhang}, \citenamefont {Chen},
  \citenamefont {He}, \citenamefont {Dong}, \citenamefont {Han}, \citenamefont
  {Li},\ and\ \citenamefont {Guo}}]{Tao_2022}%
  \BibitemOpen
  \bibfield  {author} {\bibinfo {author} {\bibfnamefont {S.-J.}\ \bibnamefont
  {Tao}}, \bibinfo {author} {\bibfnamefont {H.-Q.}\ \bibnamefont {Zhang}},
  \bibinfo {author} {\bibfnamefont {H.}~\bibnamefont {Zhang}}, \bibinfo
  {author} {\bibfnamefont {Z.}~\bibnamefont {Chen}}, \bibinfo {author}
  {\bibfnamefont {R.}~\bibnamefont {He}}, \bibinfo {author} {\bibfnamefont
  {S.-J.}\ \bibnamefont {Dong}}, \bibinfo {author} {\bibfnamefont {Y.-J.}\
  \bibnamefont {Han}}, \bibinfo {author} {\bibfnamefont {C.-F.}\ \bibnamefont
  {Li}}, \ and\ \bibinfo {author} {\bibfnamefont {G.-C.}\ \bibnamefont {Guo}},\
  }\href {\doibase 10.1103/PhysRevA.106.032403} {\bibfield  {journal} {\bibinfo
   {journal} {Phys. Rev. A}\ }\textbf {\bibinfo {volume} {106}},\ \bibinfo
  {pages} {032403} (\bibinfo {year} {2022})}\BibitemShut {NoStop}%
\bibitem [{\citenamefont {Mendon\ifmmode~\mbox{\c{c}}\else \c{c}\fi{}a}\ and\
  \citenamefont {Jachymski}(2023)}]{Mendonca_2023}%
  \BibitemOpen
  \bibfield  {author} {\bibinfo {author} {\bibfnamefont {J.~a.~P.}\
  \bibnamefont {Mendon\ifmmode~\mbox{\c{c}}\else \c{c}\fi{}a}}\ and\ \bibinfo
  {author} {\bibfnamefont {K.}~\bibnamefont {Jachymski}},\ }\href {\doibase
  10.1103/PhysRevA.107.032808} {\bibfield  {journal} {\bibinfo  {journal}
  {Phys. Rev. A}\ }\textbf {\bibinfo {volume} {107}},\ \bibinfo {pages}
  {032808} (\bibinfo {year} {2023})}\BibitemShut {NoStop}%
\bibitem [{\citenamefont {Slobodeniuk}\ \emph {et~al.}(2022)\citenamefont
  {Slobodeniuk}, \citenamefont {Novotn{\'{y}}},\ and\ \citenamefont
  {Filip}}]{Slobodeniuk_2022}%
  \BibitemOpen
  \bibfield  {author} {\bibinfo {author} {\bibfnamefont {A.}~\bibnamefont
  {Slobodeniuk}}, \bibinfo {author} {\bibfnamefont {T.}~\bibnamefont
  {Novotn{\'{y}}}}, \ and\ \bibinfo {author} {\bibfnamefont {R.}~\bibnamefont
  {Filip}},\ }\href {\doibase 10.22331/q-2022-04-15-689} {\bibfield  {journal}
  {\bibinfo  {journal} {Quantum}\ }\textbf {\bibinfo {volume} {6}},\ \bibinfo
  {pages} {689} (\bibinfo {year} {2022})}\BibitemShut {NoStop}%
\bibitem [{\citenamefont {Slobodeniuk}\ \emph {et~al.}(2023)\citenamefont
  {Slobodeniuk}, \citenamefont {Novotný},\ and\ \citenamefont
  {Filip}}]{Slobodeniuk_2023}%
  \BibitemOpen
  \bibfield  {author} {\bibinfo {author} {\bibfnamefont {A.}~\bibnamefont
  {Slobodeniuk}}, \bibinfo {author} {\bibfnamefont {T.}~\bibnamefont
  {Novotný}}, \ and\ \bibinfo {author} {\bibfnamefont {R.}~\bibnamefont
  {Filip}},\ }\href@noop {} {\  (\bibinfo {year} {2023})},\ \Eprint
  {http://arxiv.org/abs/2303.07795} {arXiv:2303.07795 [quant-ph]} \BibitemShut
  {NoStop}%
\bibitem [{\citenamefont {Becker}\ \emph {et~al.}(2021)\citenamefont {Becker},
  \citenamefont {Wu},\ and\ \citenamefont {Eckardt}}]{Becker_2021}%
  \BibitemOpen
  \bibfield  {author} {\bibinfo {author} {\bibfnamefont {T.}~\bibnamefont
  {Becker}}, \bibinfo {author} {\bibfnamefont {L.-N.}\ \bibnamefont {Wu}}, \
  and\ \bibinfo {author} {\bibfnamefont {A.}~\bibnamefont {Eckardt}},\ }\href
  {\doibase 10.1103/PhysRevE.104.014110} {\bibfield  {journal} {\bibinfo
  {journal} {Phys. Rev. E}\ }\textbf {\bibinfo {volume} {104}},\ \bibinfo
  {pages} {014110} (\bibinfo {year} {2021})}\BibitemShut {NoStop}%
\bibitem [{\citenamefont {Trushechkin}(2021)}]{Trushechkin_2021}%
  \BibitemOpen
  \bibfield  {author} {\bibinfo {author} {\bibfnamefont {A.}~\bibnamefont
  {Trushechkin}},\ }\href {\doibase 10.1103/PhysRevA.103.062226} {\bibfield
  {journal} {\bibinfo  {journal} {Phys. Rev. A}\ }\textbf {\bibinfo {volume}
  {103}},\ \bibinfo {pages} {062226} (\bibinfo {year} {2021})}\BibitemShut
  {NoStop}%
\bibitem [{\citenamefont {Davidovi{\'{c}}}(2020)}]{Davidovic_2020}%
  \BibitemOpen
  \bibfield  {author} {\bibinfo {author} {\bibfnamefont {D.}~\bibnamefont
  {Davidovi{\'{c}}}},\ }\href {\doibase 10.22331/q-2020-09-21-326} {\bibfield
  {journal} {\bibinfo  {journal} {Quantum}\ }\textbf {\bibinfo {volume} {4}},\
  \bibinfo {pages} {326} (\bibinfo {year} {2020})}\BibitemShut {NoStop}%
\bibitem [{\citenamefont {Nathan}\ and\ \citenamefont
  {Rudner}(2020)}]{Nathan_2020}%
  \BibitemOpen
  \bibfield  {author} {\bibinfo {author} {\bibfnamefont {F.}~\bibnamefont
  {Nathan}}\ and\ \bibinfo {author} {\bibfnamefont {M.~S.}\ \bibnamefont
  {Rudner}},\ }\href {\doibase 10.1103/PhysRevB.102.115109} {\bibfield
  {journal} {\bibinfo  {journal} {Phys. Rev. B}\ }\textbf {\bibinfo {volume}
  {102}},\ \bibinfo {pages} {115109} (\bibinfo {year} {2020})}\BibitemShut
  {NoStop}%
\bibitem [{\citenamefont {Mozgunov}\ and\ \citenamefont
  {Lidar}(2020)}]{Mozgunov_2020}%
  \BibitemOpen
  \bibfield  {author} {\bibinfo {author} {\bibfnamefont {E.}~\bibnamefont
  {Mozgunov}}\ and\ \bibinfo {author} {\bibfnamefont {D.}~\bibnamefont
  {Lidar}},\ }\href {\doibase 10.22331/q-2020-02-06-227} {\bibfield  {journal}
  {\bibinfo  {journal} {Quantum}\ }\textbf {\bibinfo {volume} {4}},\ \bibinfo
  {pages} {227} (\bibinfo {year} {2020})}\BibitemShut {NoStop}%
\bibitem [{\citenamefont {Becker}\ \emph {et~al.}(2022)\citenamefont {Becker},
  \citenamefont {Schnell},\ and\ \citenamefont {Thingna}}]{Becker_2022}%
  \BibitemOpen
  \bibfield  {author} {\bibinfo {author} {\bibfnamefont {T.}~\bibnamefont
  {Becker}}, \bibinfo {author} {\bibfnamefont {A.}~\bibnamefont {Schnell}}, \
  and\ \bibinfo {author} {\bibfnamefont {J.}~\bibnamefont {Thingna}},\ }\href
  {\doibase 10.1103/PhysRevLett.129.200403} {\bibfield  {journal} {\bibinfo
  {journal} {Phys. Rev. Lett.}\ }\textbf {\bibinfo {volume} {129}},\ \bibinfo
  {pages} {200403} (\bibinfo {year} {2022})}\BibitemShut {NoStop}%
\end{thebibliography}%

\end{document}